\documentclass[epj]{webofc}
\usepackage[varg]{txfonts}   
\wocname{EPJ Web of Conferences}
\woctitle{New Frontiers in Physics 2014}
\begin{document}
\selectlanguage{english}
\title{ Recent results from NA61/SHINE }
%
%

\author{Marek Gazdzicki\inst{1,2}\fnsep\thanks{\email{marek@cern.ch} }
        for the NA61/SHINE Collaboration  
}

\institute{ Goethe--University, Frankfurt, Germany 
\and
            Jan Kochanowski University, Kielce, Poland
          }

\abstract{%
   This paper briefly presents the NA61/SHINE facility at the CERN SPS 
   and its measurements motivated by physics of strong interactions,
   neutrinos and cosmic rays. 
}
\maketitle
\section{Introduction}
\label{sec:Introduction}
NA61/SHINE is a multi-purpose
experimental facility to study hadron production in hadron-proton,
hadron-nucleus and nucleus-nucleus collisions at
the CERN Super Proton Synchrotron.
Measurements motivated by physics of strong interactions,
neutrinos and cosmic rays have been performed up to now.
The first physics data with secondary hadron (protons, pions and kaons) 
beams were recorded 
in 2009 and with nuclear beams (secondary $^7$Be beams) in 2011.

This contribution briefly presents the NA61/SHINE 
facility~\cite{Abgrall:2014xwa} in
Sec.~\ref{sec:Facility} and 
recent results~\cite{Abgrall:2014sr} on the three physics programs  in
Secs.~\ref{sec:Strong_interactions}, \ref{sec:Neutrinos},
and \ref{sec:Cosmic_rays}.
The contribution consists of copies of the slides presented at
the conference {\it New Frontiers in Physics 2014} included as
figures with captions.

\begin{center}
\vspace{1.5cm}
\includegraphics[width=6cm]{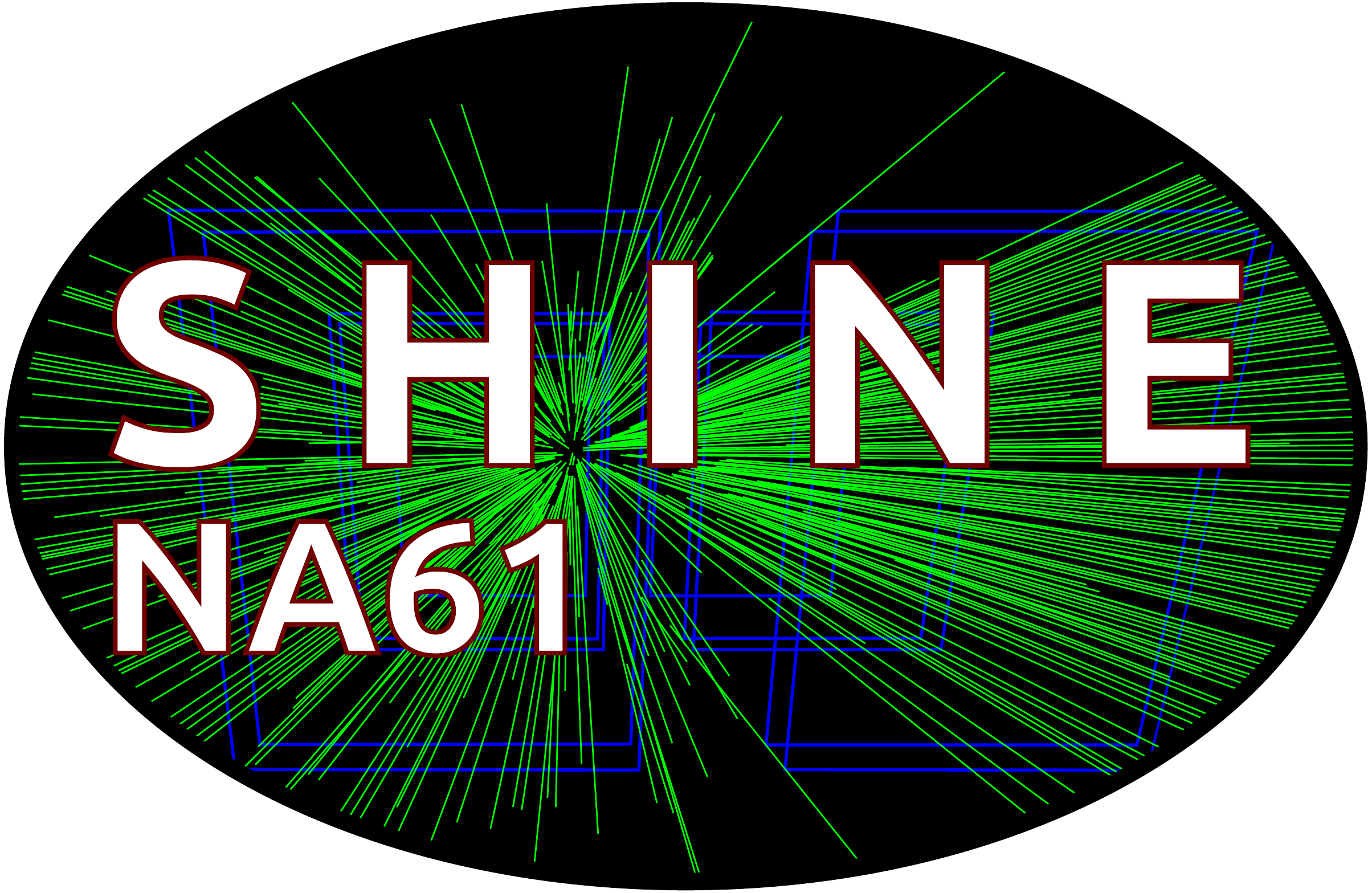}\\
\vspace{1.5cm}
\end{center}

\newpage

\section{Facility}
\label{sec:Facility}
\begin{figure}[h]
\centering
\includegraphics[width=7cm,clip]{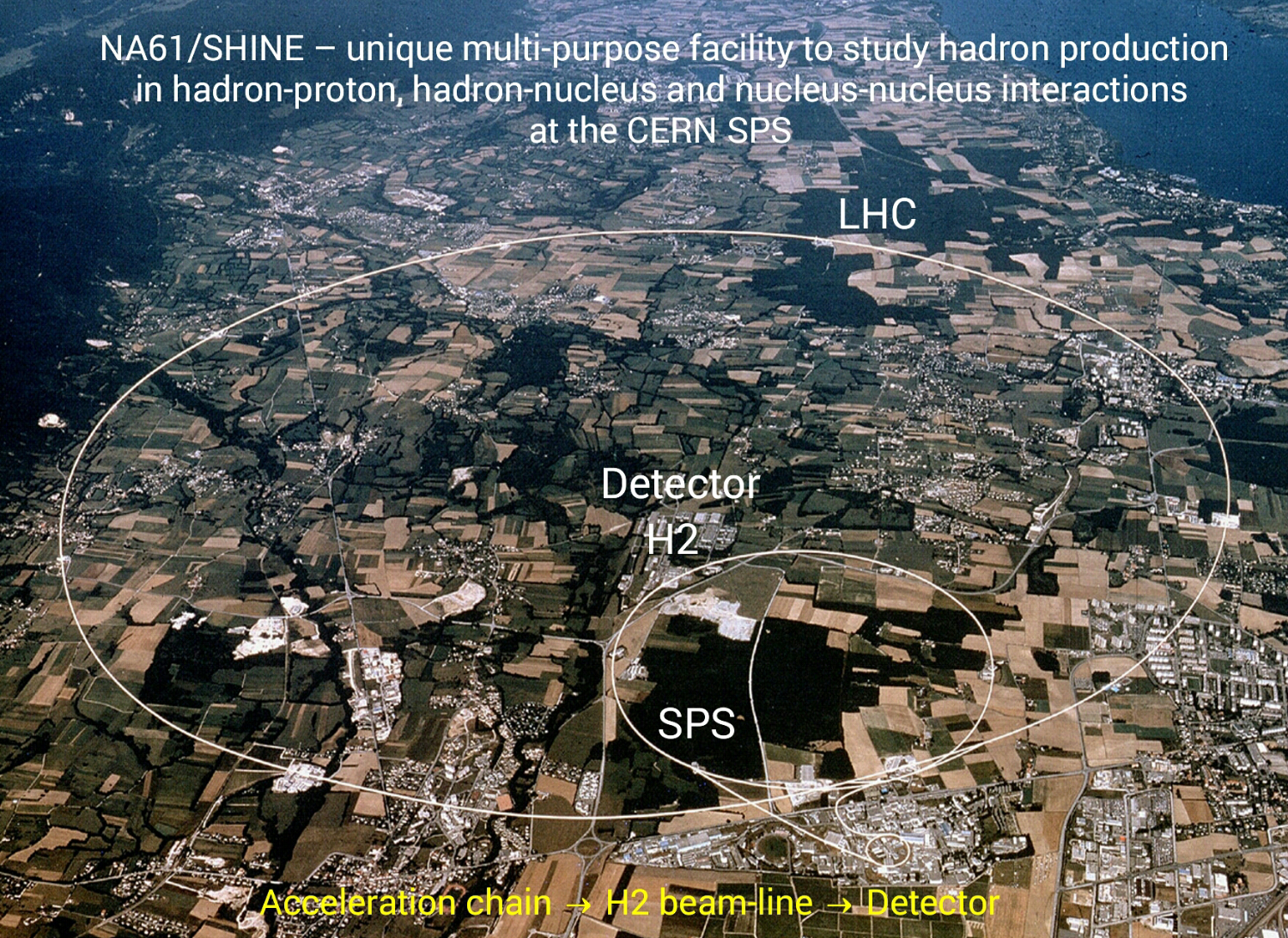}
\caption{
The NA61/SHINE facility is located at the European Organization for
Nuclear Research, CERN, on the Franco-Swiss border near Geneva.
The acceleration chain delivers  proton and nuclear beams
to the T2 target located in the North Area target cavern at the
beginning of the H2 beam-line. Then the beams are 
either directly transported to the detector or used to produce
secondary beams.
}
\label{fig:na61_facility}       
\end{figure}

\begin{figure}[h]
\centering
\includegraphics[width=9cm,clip]{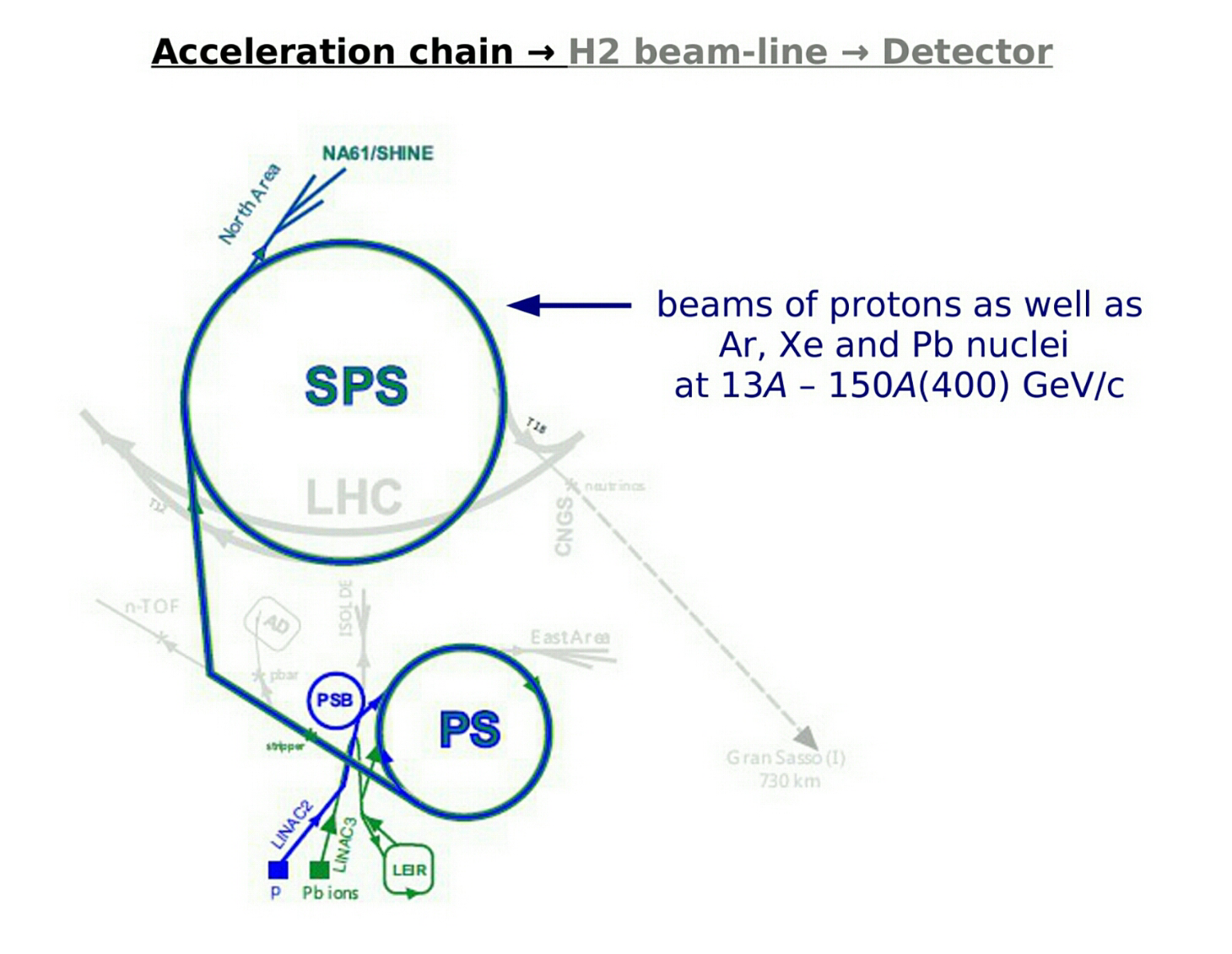}
\caption{
The proton and ion acceleration chains use different particle sources,
linear accelerators (LINAC2 and LINAC3) and circular accumulator 
machines (PSB and LEIR).
They share the Proton Synchrotron (PS) and the Super Proton Synchrotron (SPS).
}
\label{fig:na61_acc}       
\end{figure}

\begin{figure}[!h]
\centering
\includegraphics[height=5cm,clip]{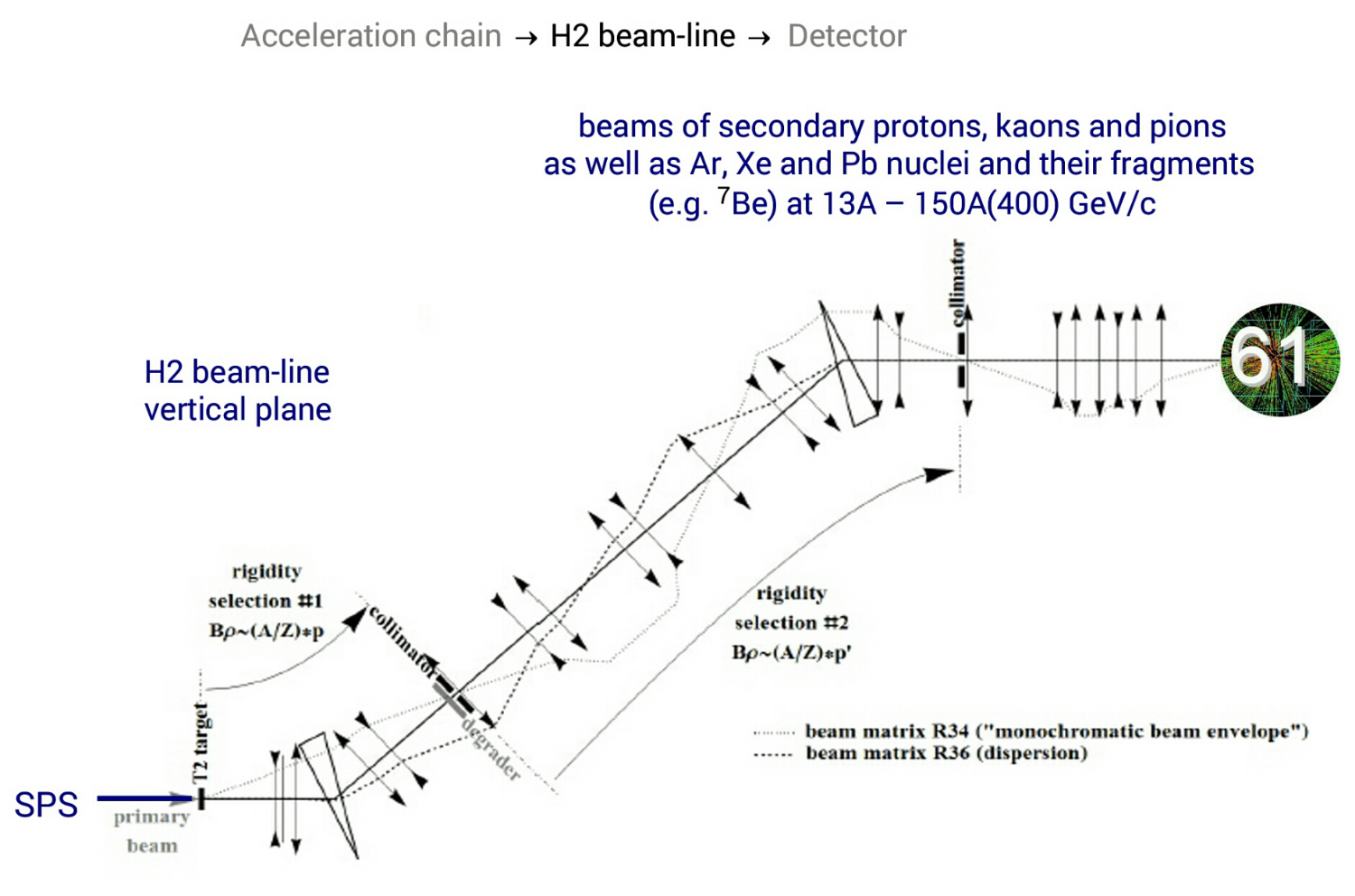}
\caption{
The beam extracted from the SPS is transported over about 1~km by
bending and focusing magnets and then split into three parts
each one directed towards a  target where secondary beams can
be created. The H2 beam-line emerges from the T2 primary
target and is able to transport momentum selected secondary particles
to the Experimental Hall North 1 where the detector
is located.
The physics target is at 535~m distance from the T2 target.
The H2 beam-line can transport charged particles in a wide range of
momenta from $\sim9$~GeV/c up to the top SPS energy of 400~GeV/c.
Alternatively it can transport a primary beam of protons or nuclei.
The momentum selection is done in the vertical plane.
The beam-line basically
consists of two long spectrometers.
}
\label{fig:na61_h2}       
\end{figure}

\begin{figure}[!h]
\centering
\includegraphics[height=6cm,clip]{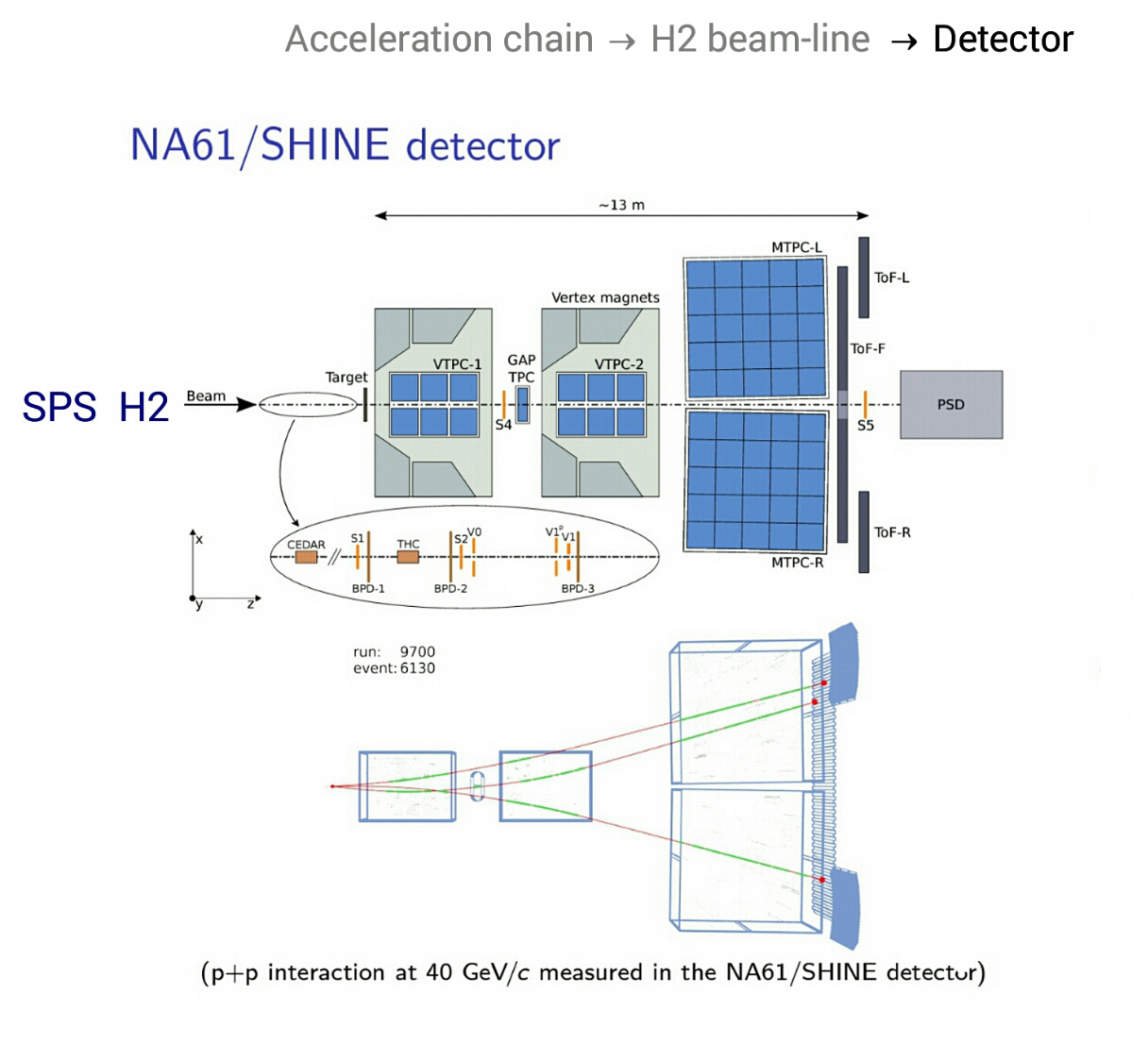}
\caption{
The NA61/SHINE detector
consists of a large acceptance hadron spectrometer with
excellent capabilities in charged particle momentum measurements and
identification by a set of six Time Projection Chambers as well as
Time-of-Flight detectors. The high resolution forward calorimeter,
the Projectile Spectator Detector, measures energy flow around the beam
direction, which in nucleus-nucleus reactions is primarily a measure of
the number of spectator (not taking part in the interaction) nucleons and thus related to the
volume of interacting matter. 
For hadron-nucleus interactions, the collision volume is determined
by counting low momentum particles emitted from the nuclear target
with the LMPD detector (a small TPC) surrounding the target.
An array of beam detectors identifies beam particles, secondary
hadrons and nuclei as well as primary nuclei, and measures precisely
their trajectories.
}
\label{fig:na61_detector}       
\end{figure}

\clearpage
\newpage

\section{Strong interactions}
\label{sec:Strong_interactions}
\begin{figure}[!h]
\centering
\includegraphics[width=6cm,clip]{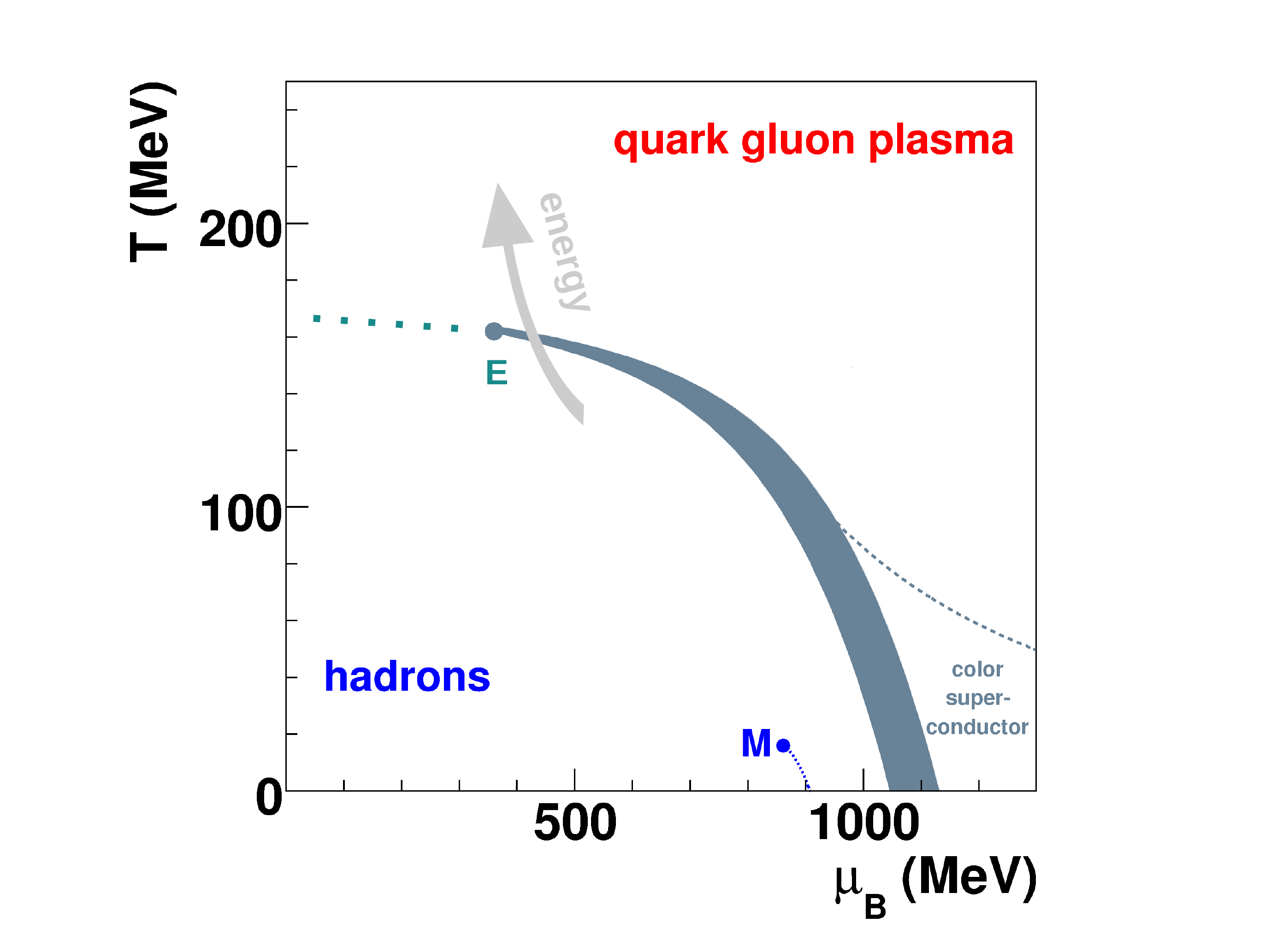}
\includegraphics[width=6cm,clip]{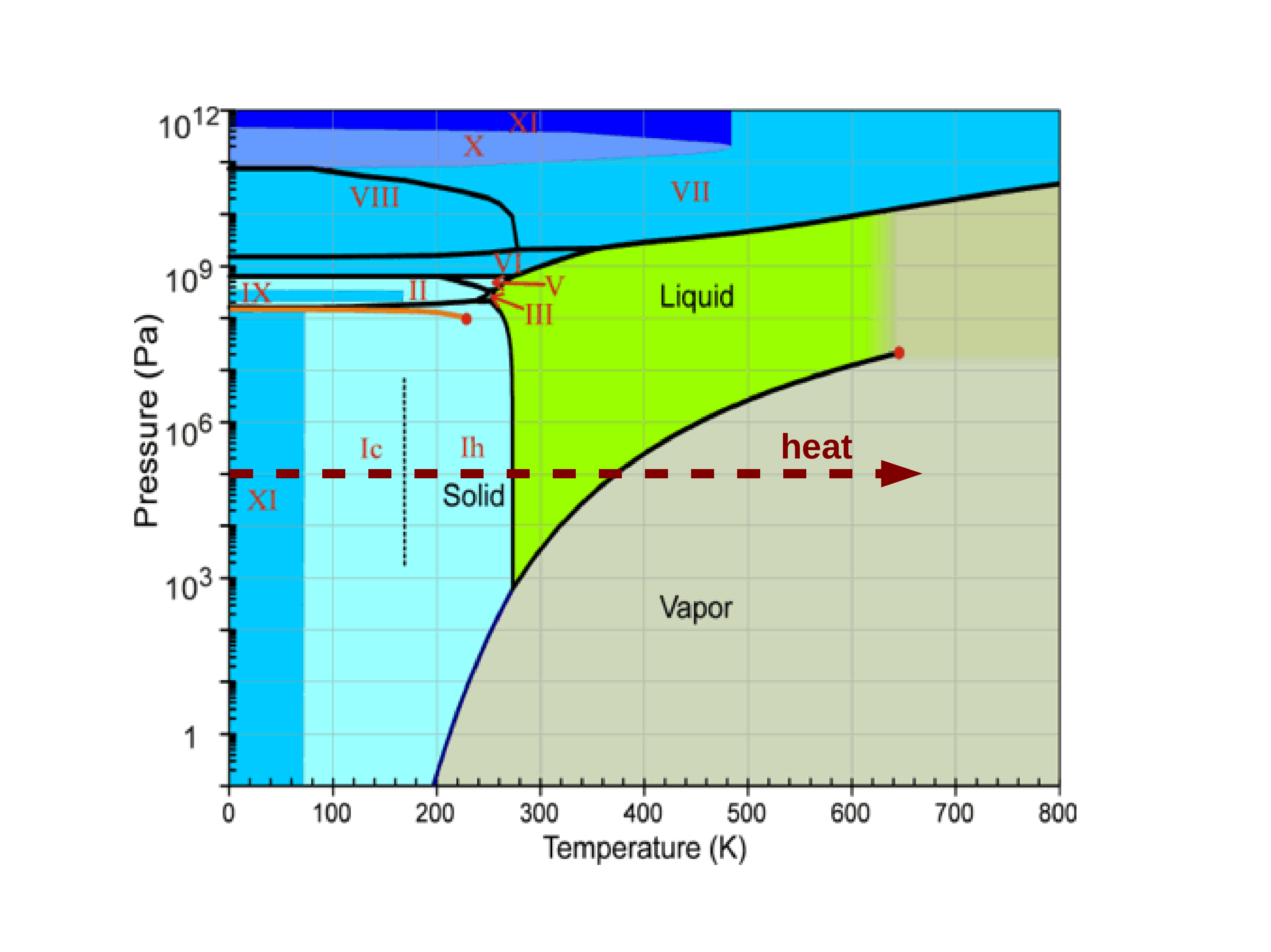}
\caption{
The NA61/SHINE measurements for physics of strong interactions are
motivated by the question what happens when strongly interacting
matter gets hotter/denser and its volume increases.
In particular, the structure of the transition domain between
matter consisting of hadrons and quark gluon plasma 
(see the left plot) is under study. 
How the signals of the onset of deconfinement observed
in central Pb+Pb collisions~collisions~\cite{Gazdzicki:2014sva} 
are modified when decreasing the size of colliding nuclei?  
Is the structure of the transition line similar to the one
of water (see the right plot) with the critical point separating the first order phase
transition from the cross-over?
}
\label{fig:na61_sim}       
\end{figure}

\begin{figure}[!h]
\centering
\includegraphics[width=8cm,clip]{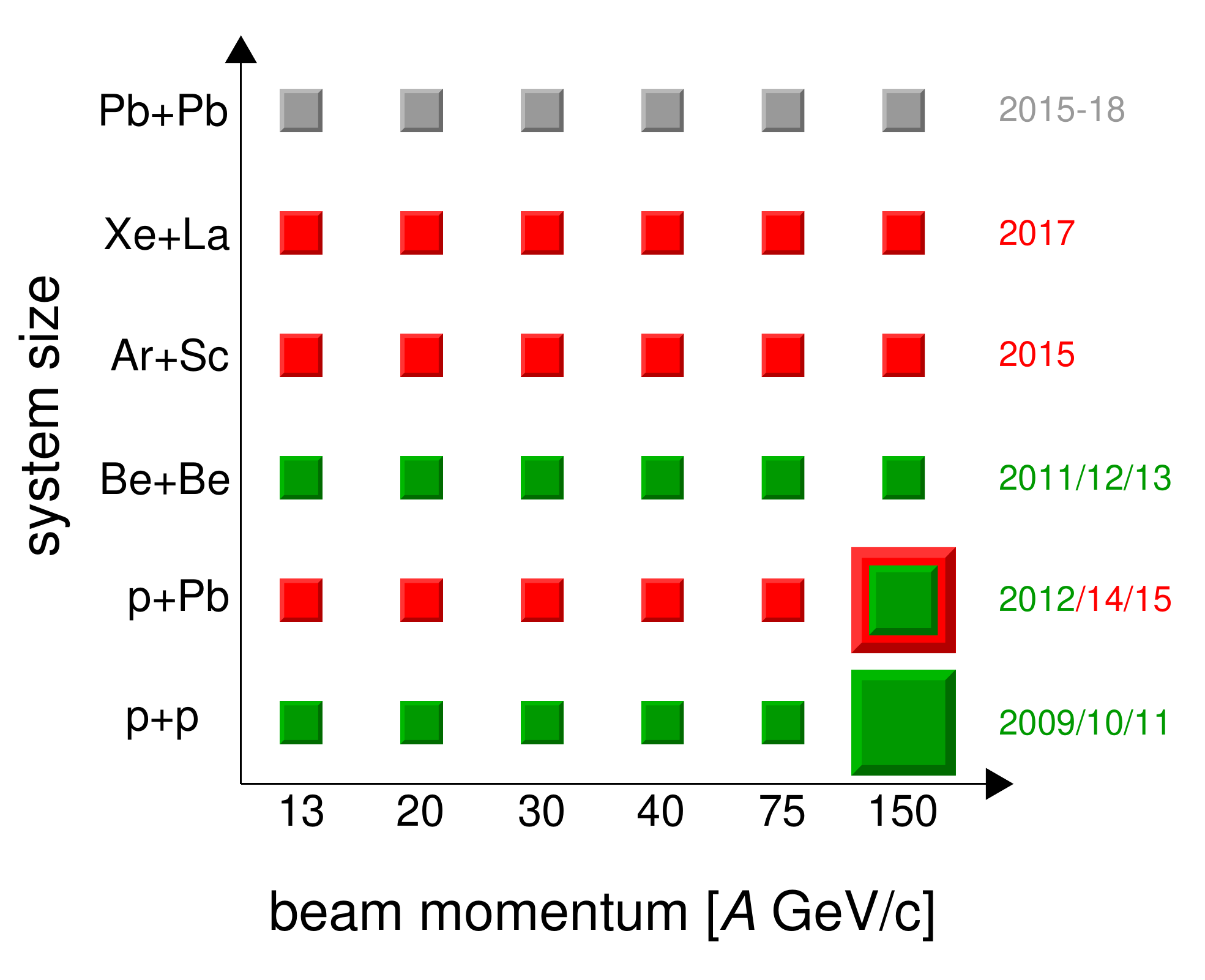}
\caption{
For the programme on strong interactions
NA61/SHINE scans in the system size and beam momentum. 
In the plot the recorded data are indicated in green,
the approved future data taking in red, whereas
the proposed extension for the period 2015--2018 in gray.
}
\label{fig:na61_schedule}       
\end{figure}

\begin{figure}[!h]
\centering
\includegraphics[width=9cm,clip]{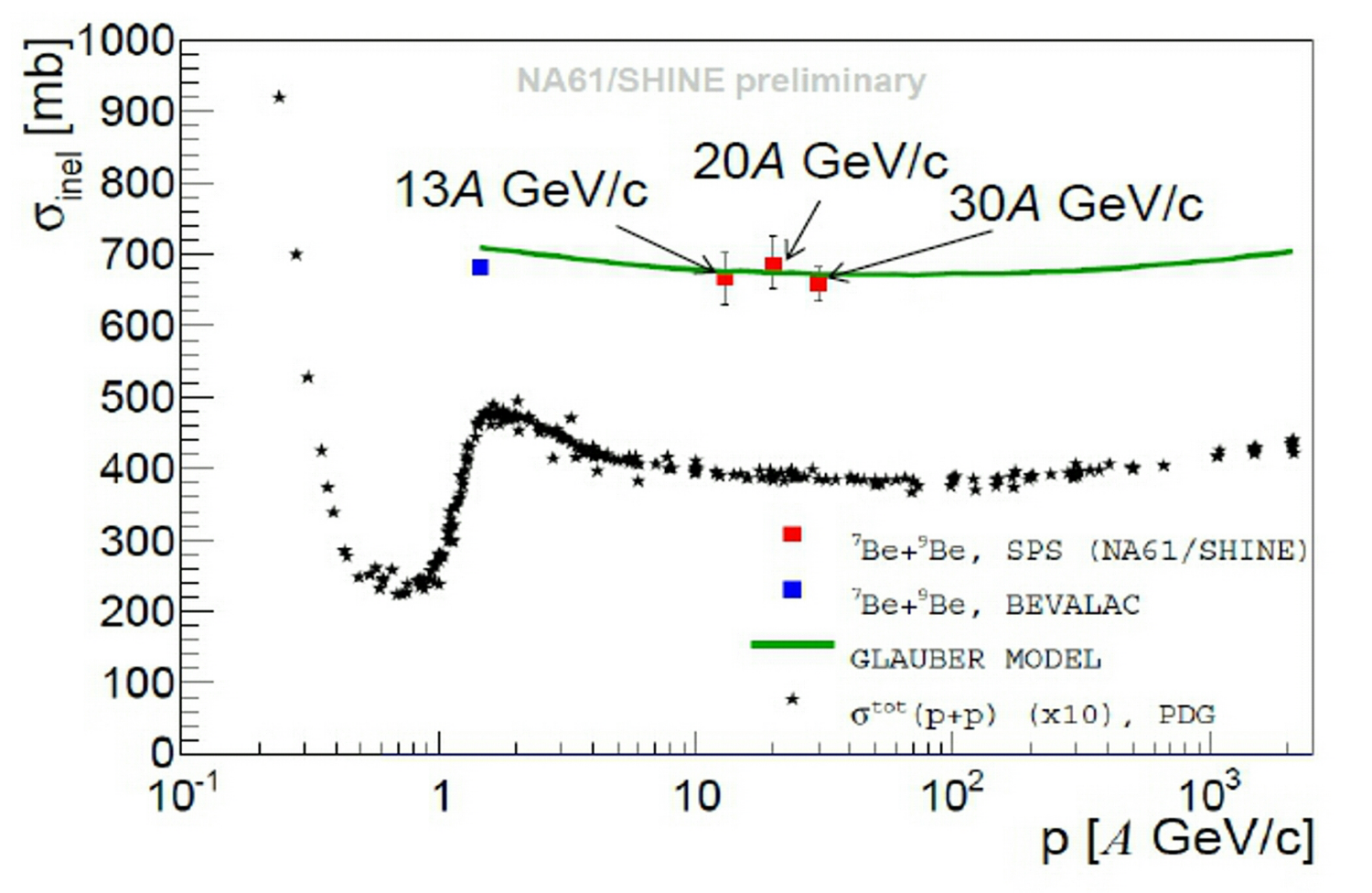}
\caption{
The inelastic cross section for $^7$Be+$^9$Be collisions is weakly dependent
on collision energy above 2$A$~GeV. 
It is well reproduced by the Glissando version of 
the Glauber model~\cite{Broniowski:2007nz}.
}
\label{fig:na61_sigmaBeBe}       
\end{figure}

\begin{figure}[!h]
\centering
\includegraphics[height=3cm,clip]{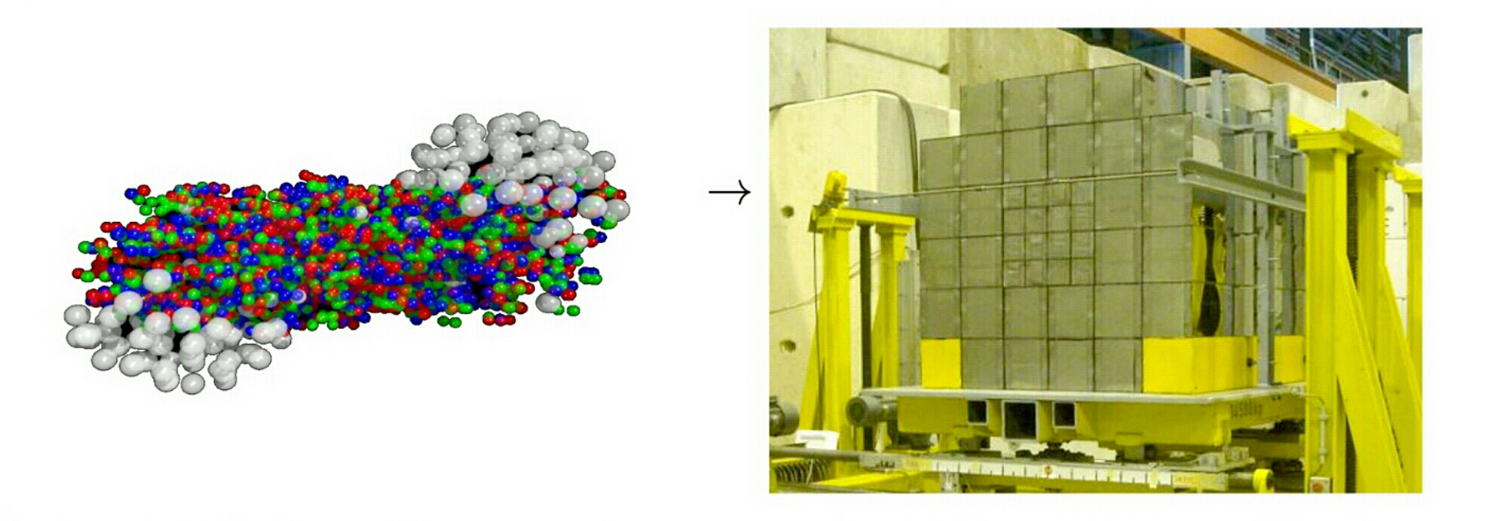}
\includegraphics[height=5cm,clip]{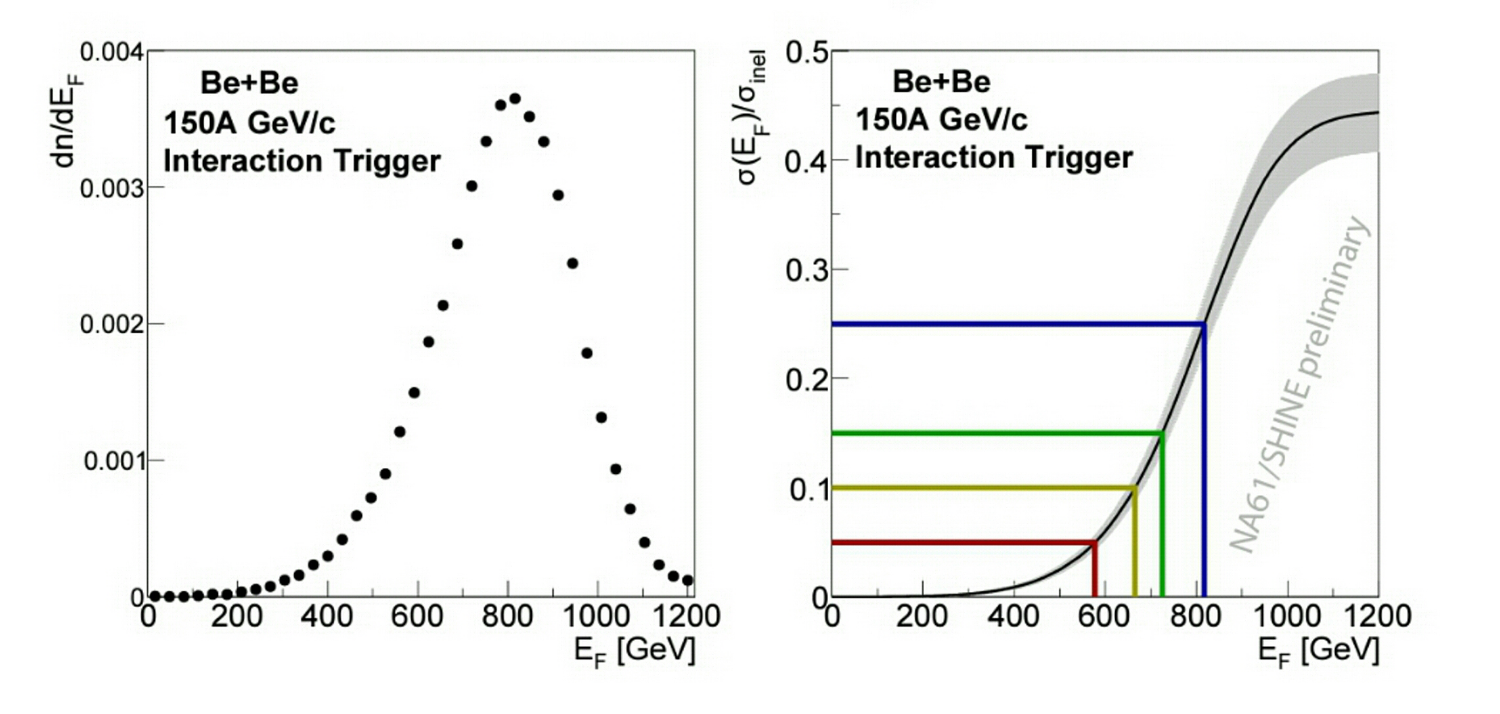}
\caption{
The selection of the reaction volume in
$^7$Be+$^9$Be collisions is performed by dividing recorded events 
into four classes according to the forward 
(mostly projectile spectator) energy, $E_F$,
measured by the Projectile Spectator Detector (see the top plots).
The first class comprises 5\% of all inelastic collisions with the smallest
$E_F$ and the next classes 5-10\%, 10-15\% and 15-20\%, respectively
(see the bottom plots).
}
\label{fig:na61_psd}       
\end{figure}

\clearpage

\begin{figure}[!h]
\centering
\includegraphics[width=12cm,clip]{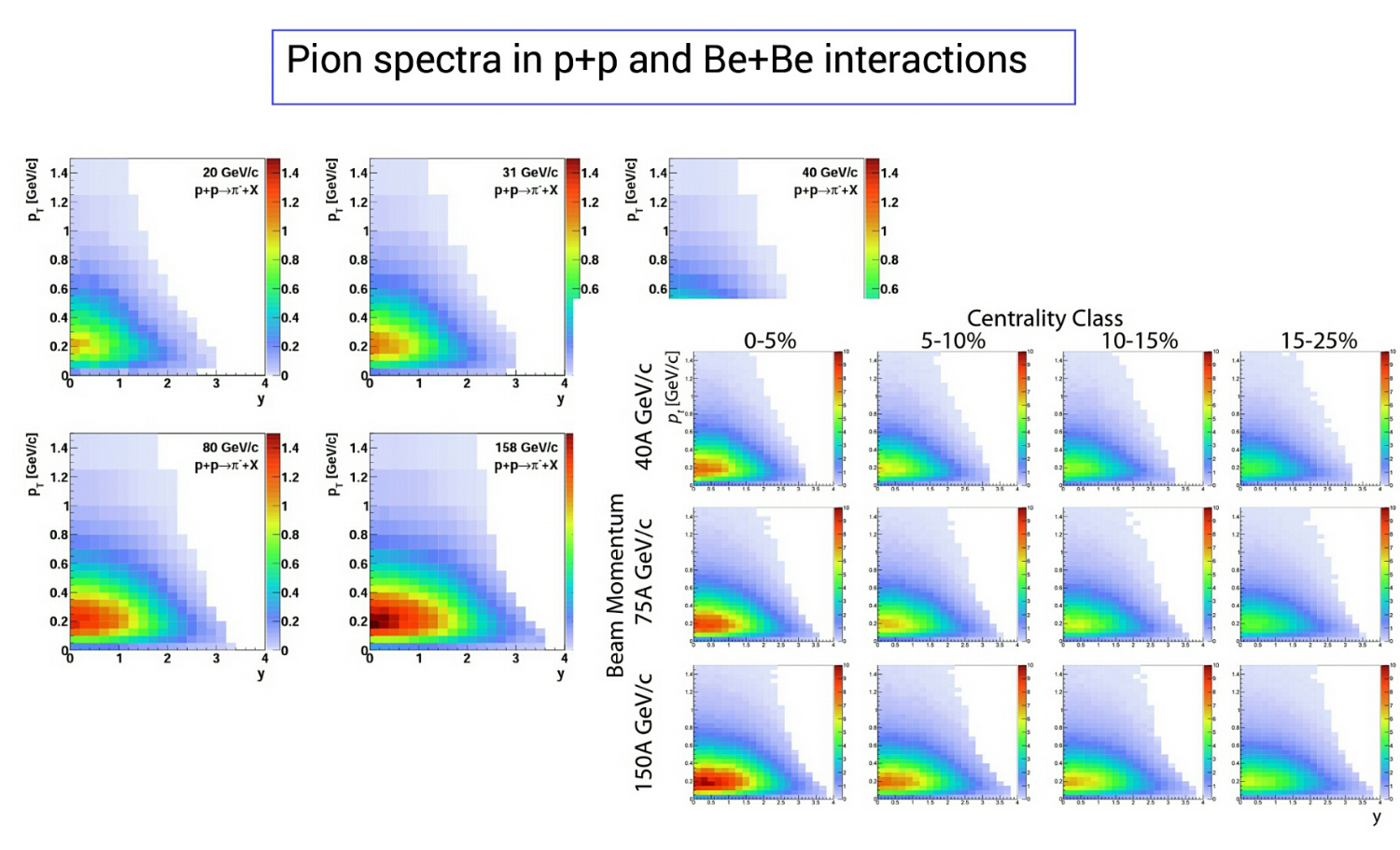}
\caption{
Negatively charged pion spectra are measured in inelastic p+p
interactions~\cite{Abgrall:2013qoa} 
and volume selected $^7$Be+$^9$Be collisions in
the SPS energy range. High statistics and large acceptance allow to obtain
two-dimensional (rapidity-transverse momentum) distributions. 
}
\label{fig:na61_pion_m}       
\end{figure}

\begin{figure}[!h]
\centering
\includegraphics[width=9cm,clip]{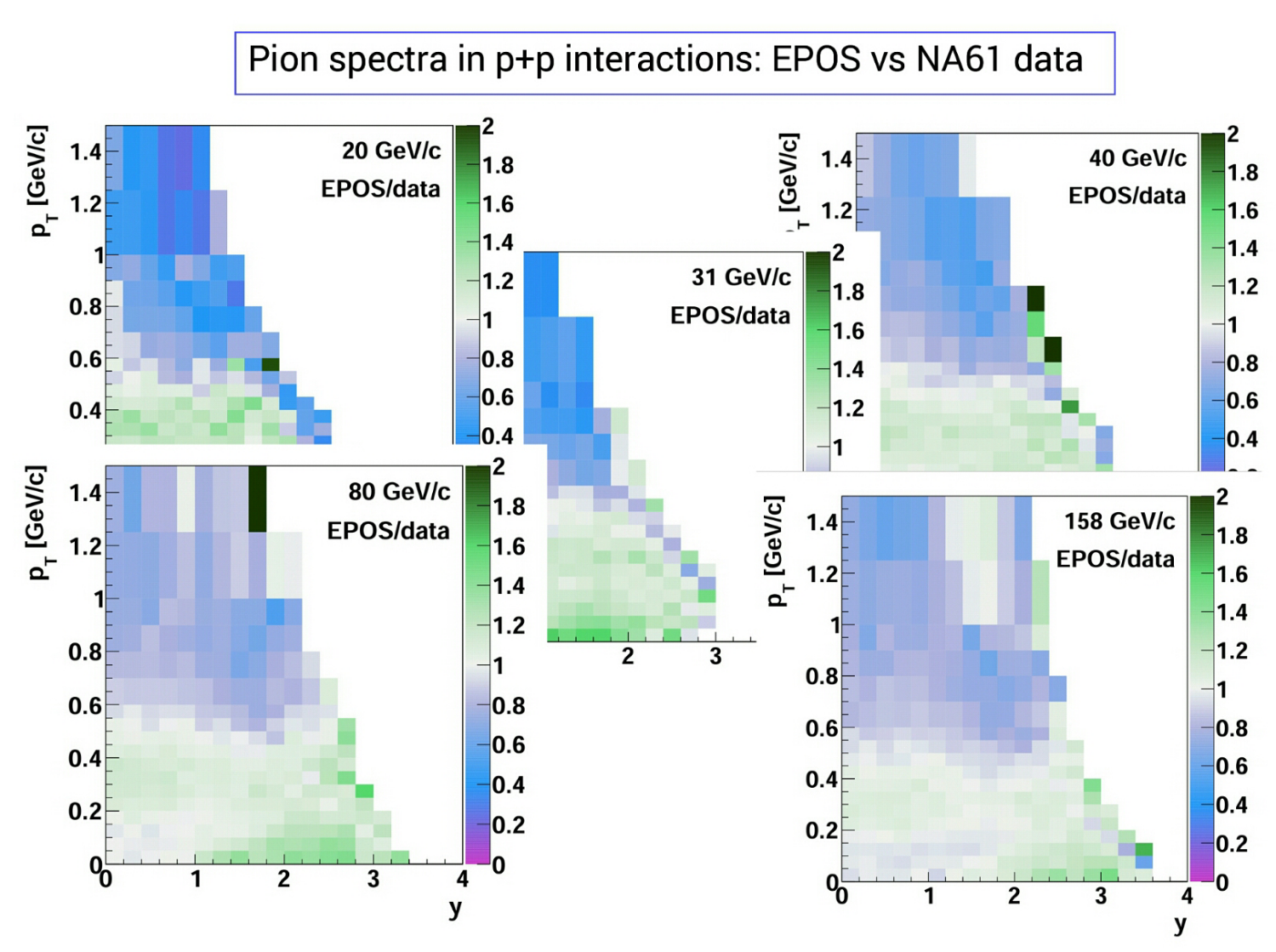}
\caption{
Negatively charged pion spectra  measured in inelastic p+p
interactions~\cite{Abgrall:2013qoa} are used to fit basic parameters
of string-resonance models~\cite{Uzhinsky:2011ir,Uzhinsky:2011qb,
Uzhinsky:2013ata,Uzhinsky:2014kxa,Vovchenko:2014ssa,Vovchenko:2014vda}
and test scaling ideas~\cite{Praszalowicz:2013uu}.
}
\label{fig:na61_pion2epos}       
\end{figure}

\clearpage

\begin{figure}[!h]
\centering
\includegraphics[width=11cm,clip]{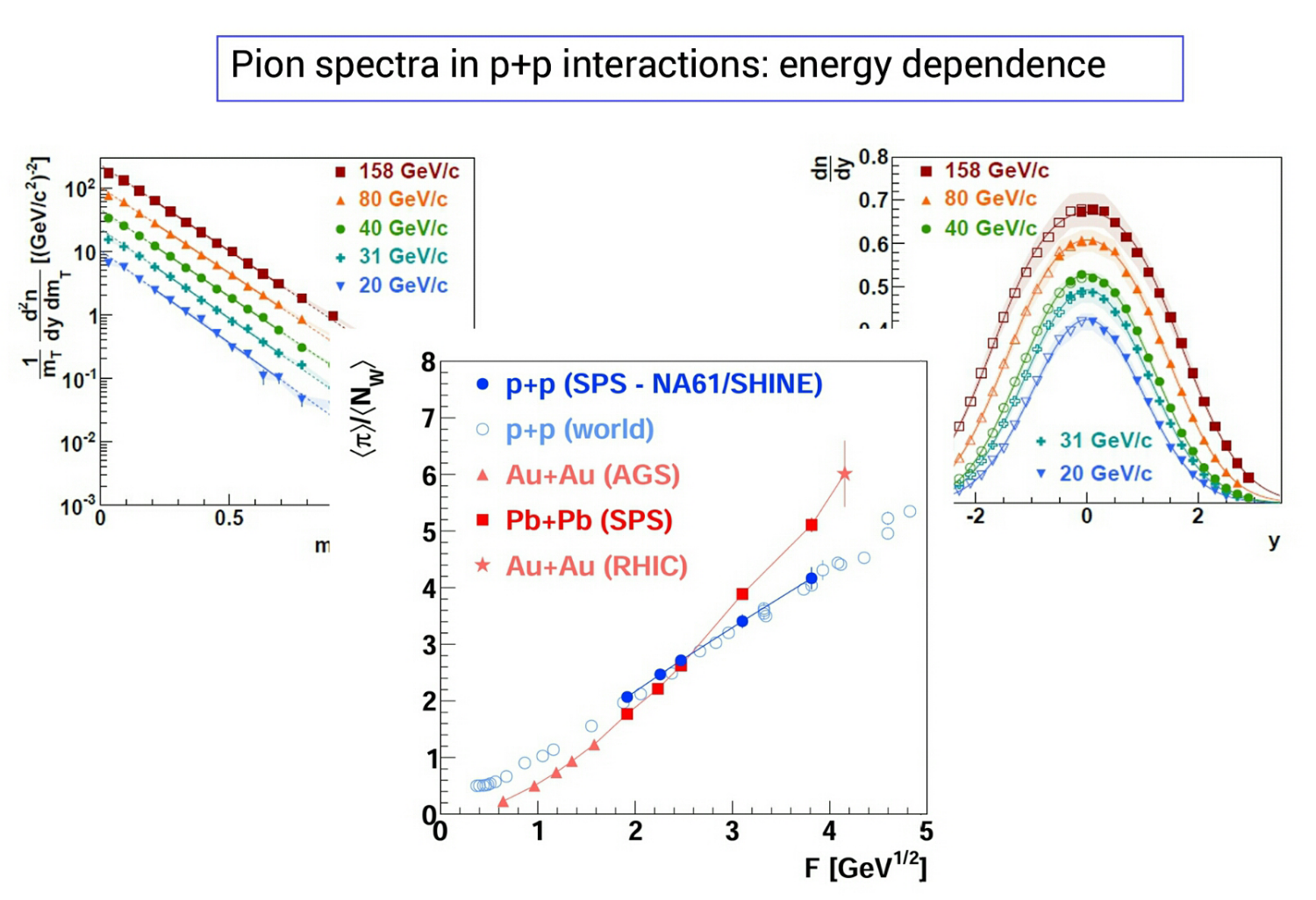}
\caption{
Collision energy dependence is studied by making slices,
projections and integrals of the two-dimensional spectra.
Examples are given for negatively charged pions:
transverse mass spectra at mid-rapidity (top, left), 
rapidity spectra (top right) and
mean pion multiplicity per wounded nucleon (bottom)~\cite{Abgrall:2013qoa}.
}
\label{fig:na61_meann.png}       
\end{figure}

\begin{figure}[!h]
\centering
\includegraphics[width=4cm,clip]{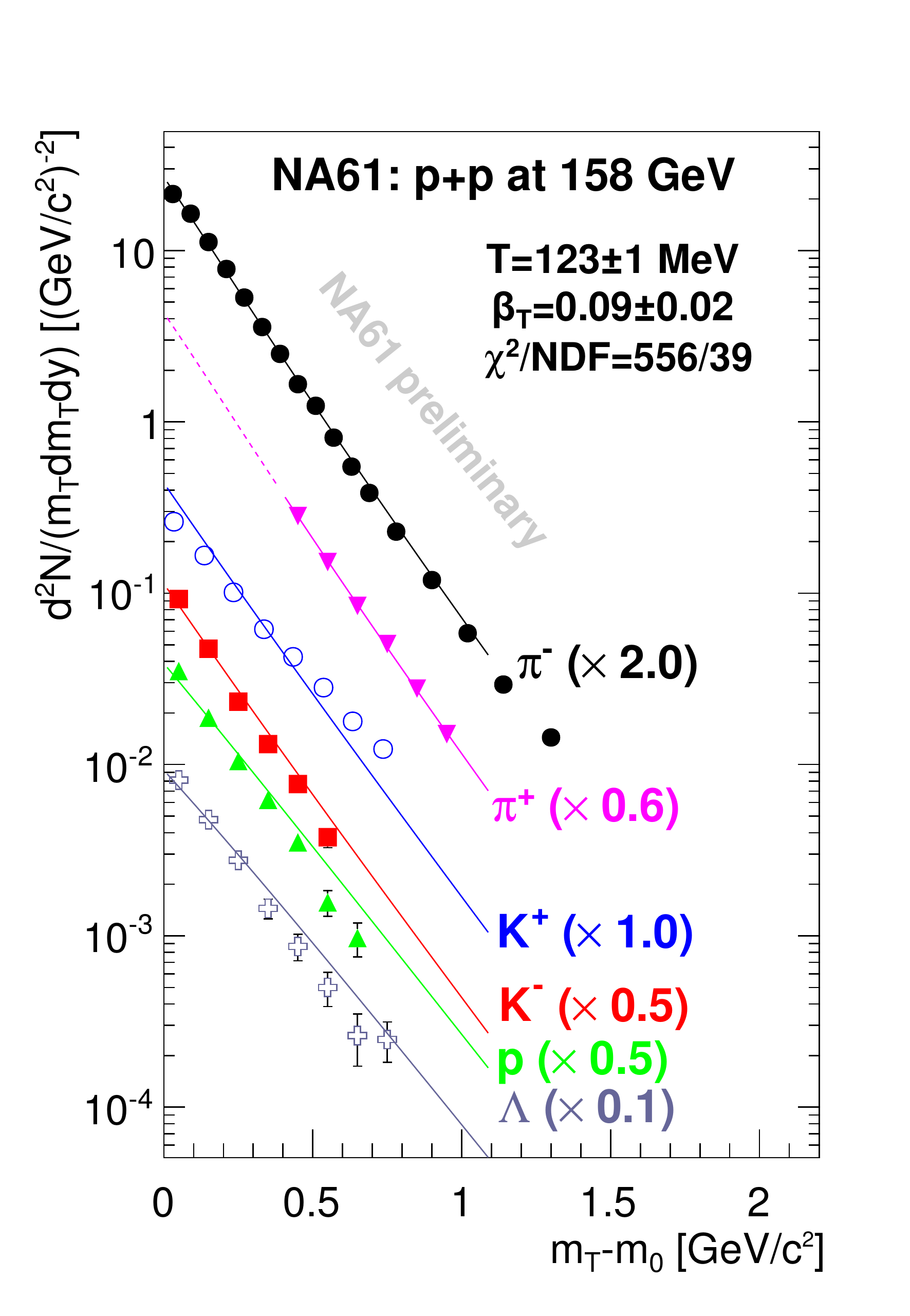}
\includegraphics[width=4cm,clip]{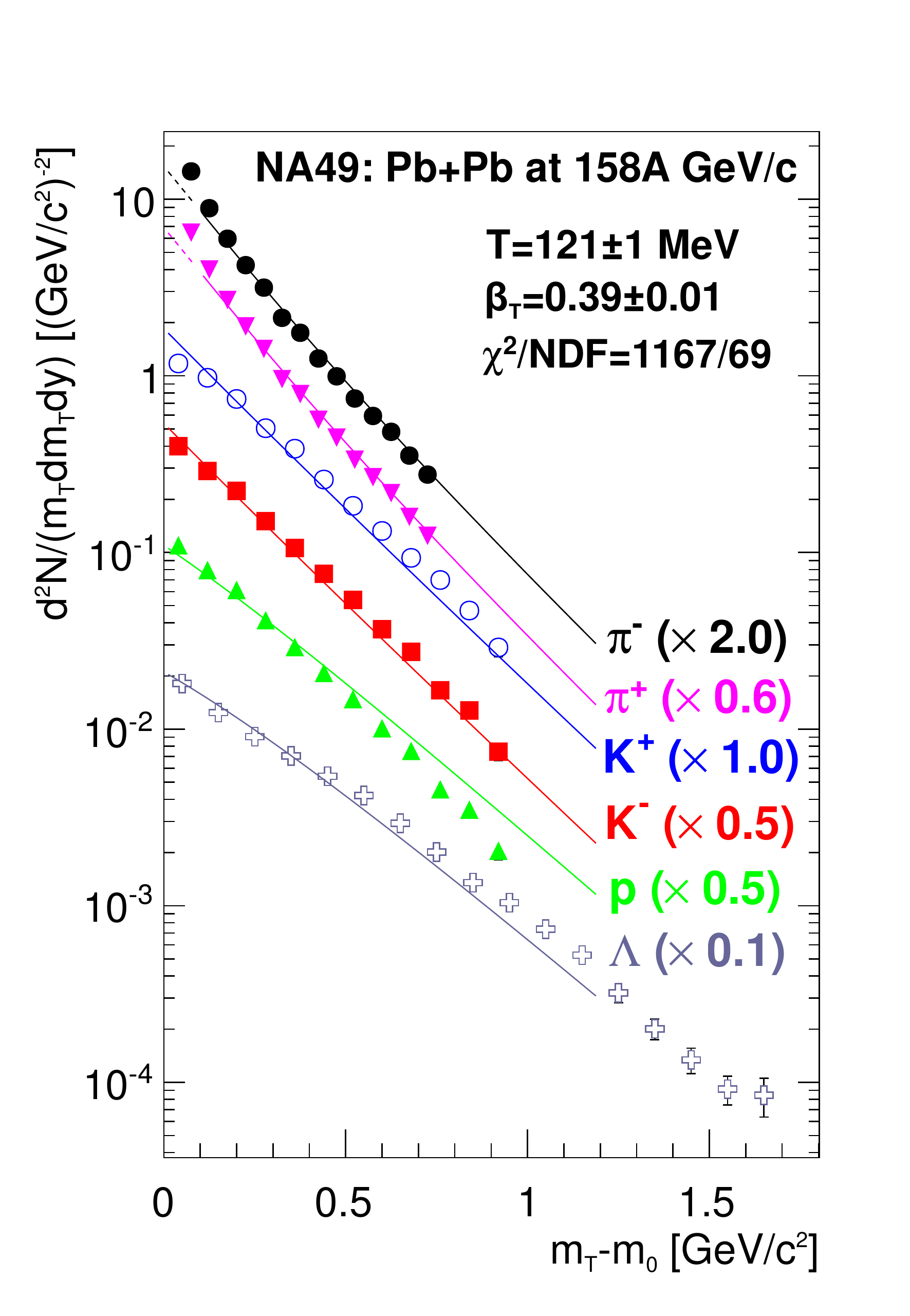}
\caption{
Mid-rapidity transverse mass spectra of
various measured particle species
are approximately exponential
in inelastic p+p interactions at 158~GeV/c. 
In central Pb+Pb collisions at 158$A$~GeV/c~\cite{Alt:2007aa, Afanasiev:2002mx}
large deviations from the exponential dependence are observed. 
It is common to attribute this finding to the transverse collective flow
which is expected to increase with increasing volume of the colliding matter
and collision energy.
}
\label{fig:na61_blast}       
\end{figure}

\begin{figure}[!h]
\centering
\includegraphics[width=9cm,clip]{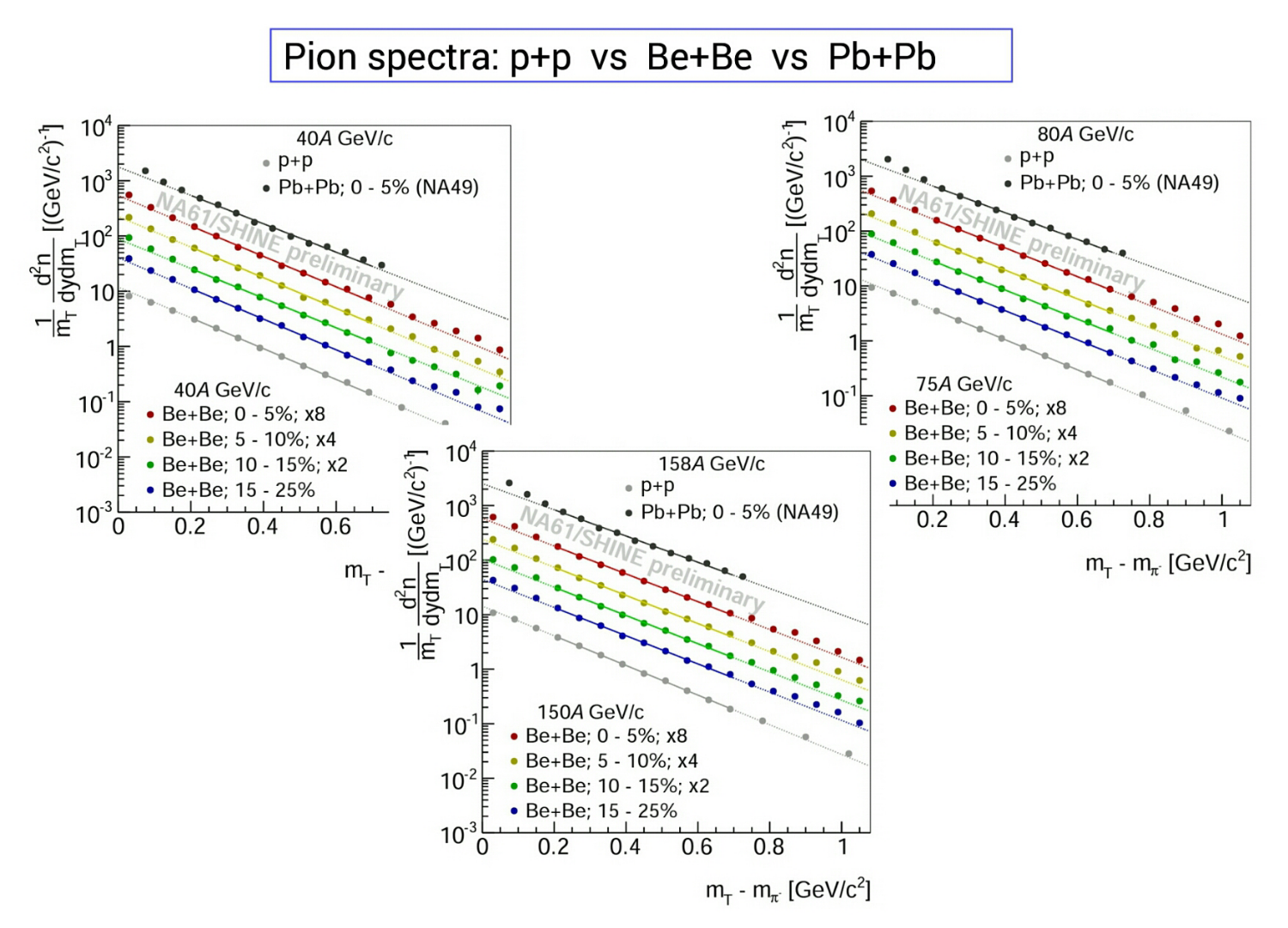}
\caption{
Mid-rapidity transverse mass spectra of negatively charged pions
measured in inelastic p+p interactions 
as well as volume selected Be+Be and Pb+Pb collisions
are compared at 40$A$, 75(80)$A$ and 150(158)$A$~GeV/c.
An exponential function with the inverse slope parameter $T$ 
is fitted in the interval
$0.2 < m_T - m_{\pi} < 0.7$~GeV (shown by lines). 
}
\label{fig:na61_mt_ppbebepbpb}       
\end{figure}

\begin{figure}[!h]
\centering
\includegraphics[width=9cm,clip]{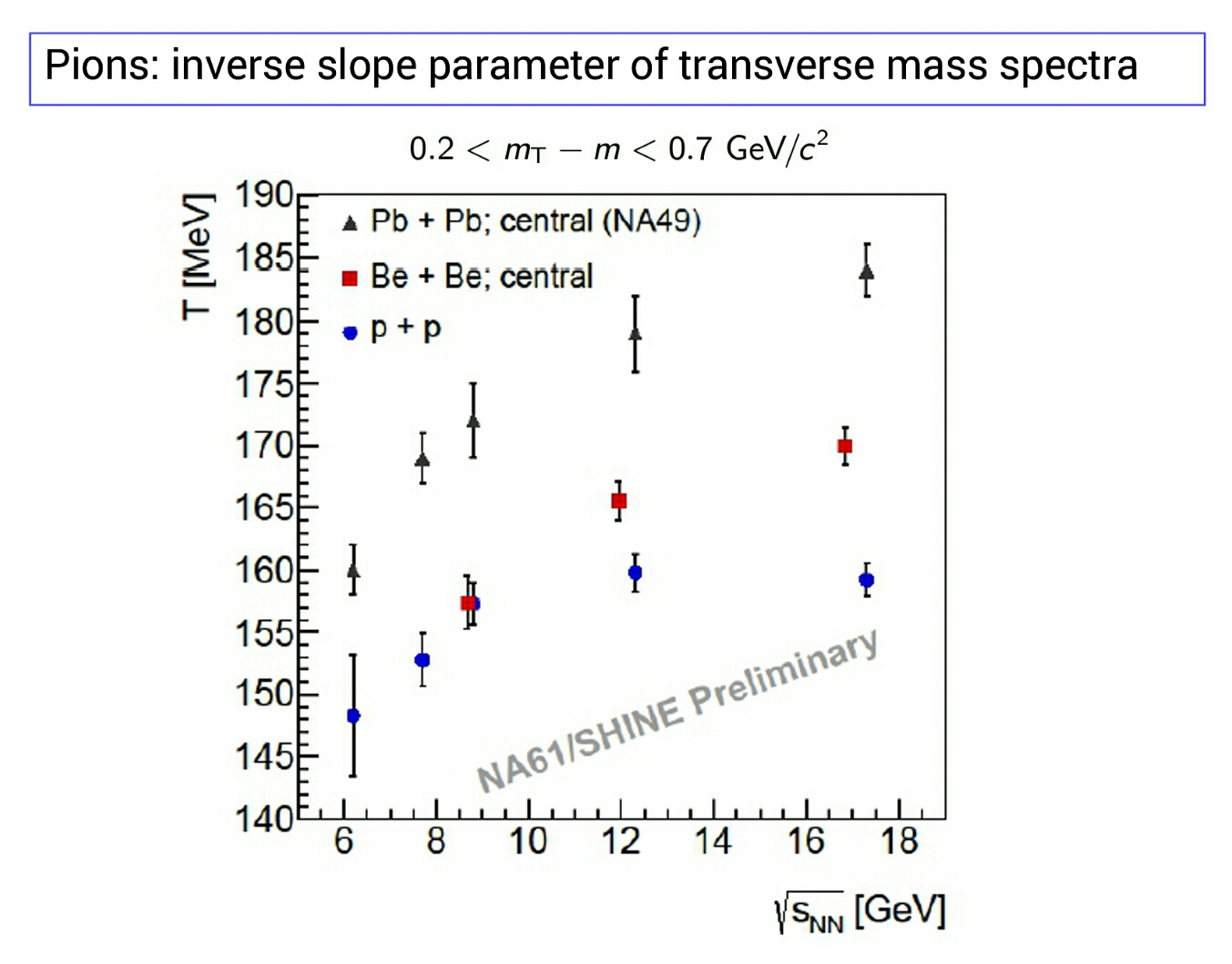}
\caption{ The $T$ parameter increases with increasing collision
energy  for all studied reactions. In large volume Pb+Pb collisions $T$ is
larger than in p+p interactions. 
The results for large volume Be+Be collisions, 
$T$(p+p) < $T$(Be+Be) < $T$(Pb+Pb) at 
75$A$ and 150$A$~GeV/c, and 
$T$(Be+Be)$\approx$$T$(p+p) at 40$A$~GeV/c,
suggest that 
the collective flow is significant in Be+Be collisions at the high SPS
beam momenta and gets reduced at 40$A$~GeV/c.
}
\label{fig:na61_onset_collectivity}       
\end{figure}

\begin{figure}[!h]
\centering
\includegraphics[width=9cm,clip]{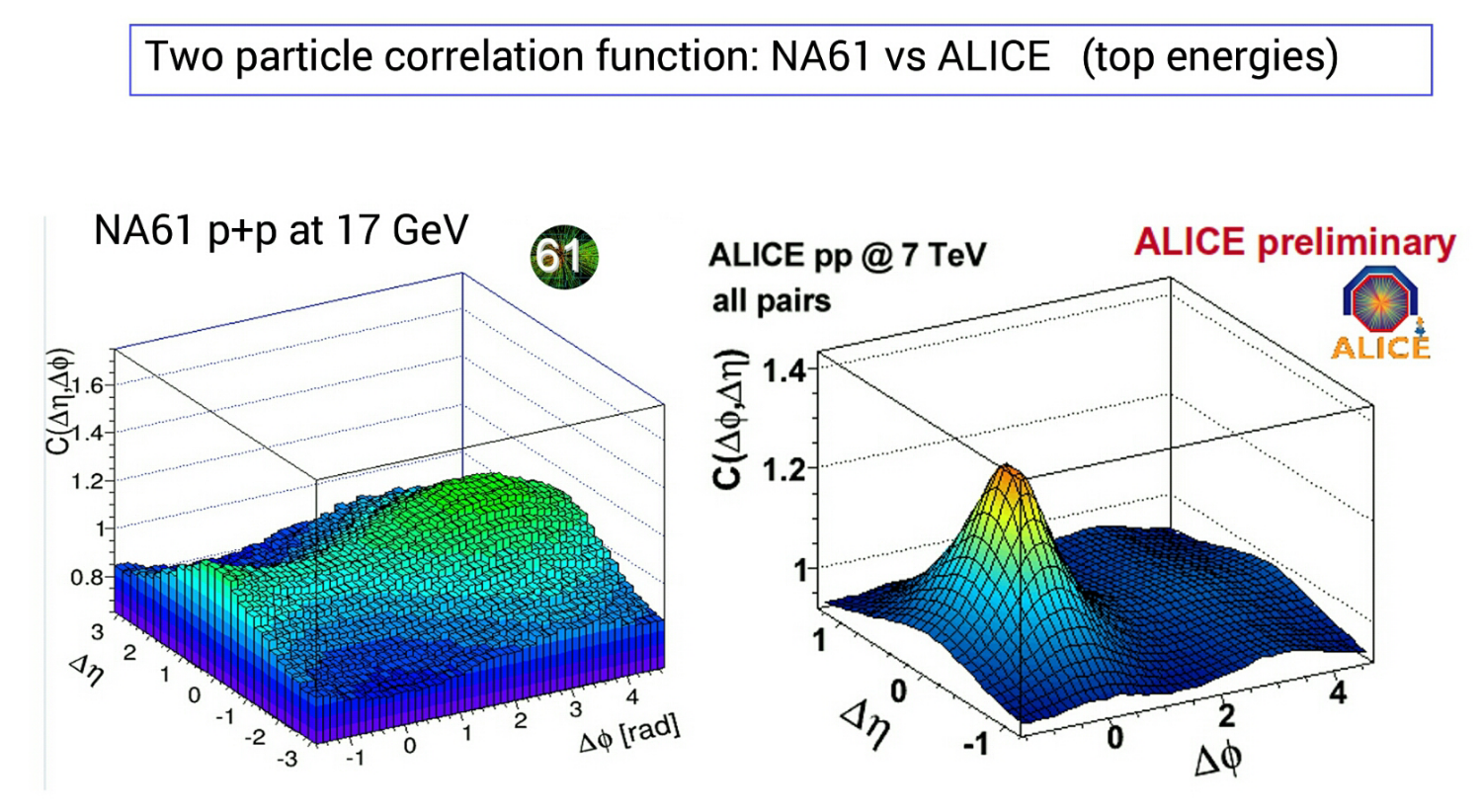}
\caption{ 
The two-particle correlation function in pseudo rapdity and
azimuthal angle in inelastic p+p interactions is very
different at the SPS and LHC energies. In particular,
the saddle at (0,0) observed at the SPS is changed
to the pronounced peak at the LHC.
}
\label{fig:na61_corr1.png}       
\end{figure}

\begin{figure}[!h]
\centering
\includegraphics[width=13cm,clip]{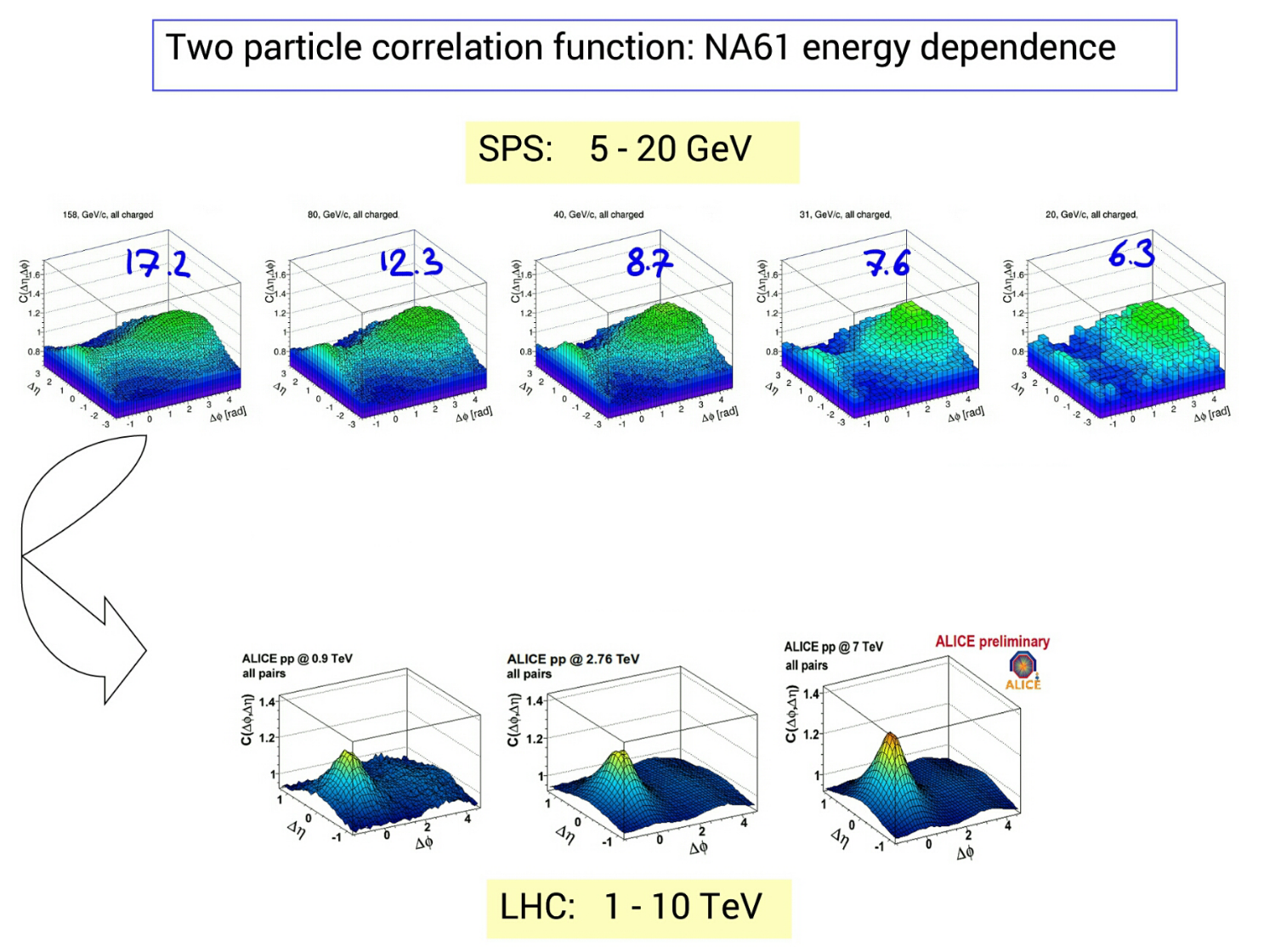}
\caption{ 
The correlation function shows monotonic evolution with
the collision energy. The height of the saddle at (0,0) 
at the SPS energies and the peak at (0,0)
at LHC increase with collision energy.
}
\label{fig:na61_corr_na61_alice}       
\end{figure}

\begin{figure}[!h]
\centering
\includegraphics[width=13cm,clip]{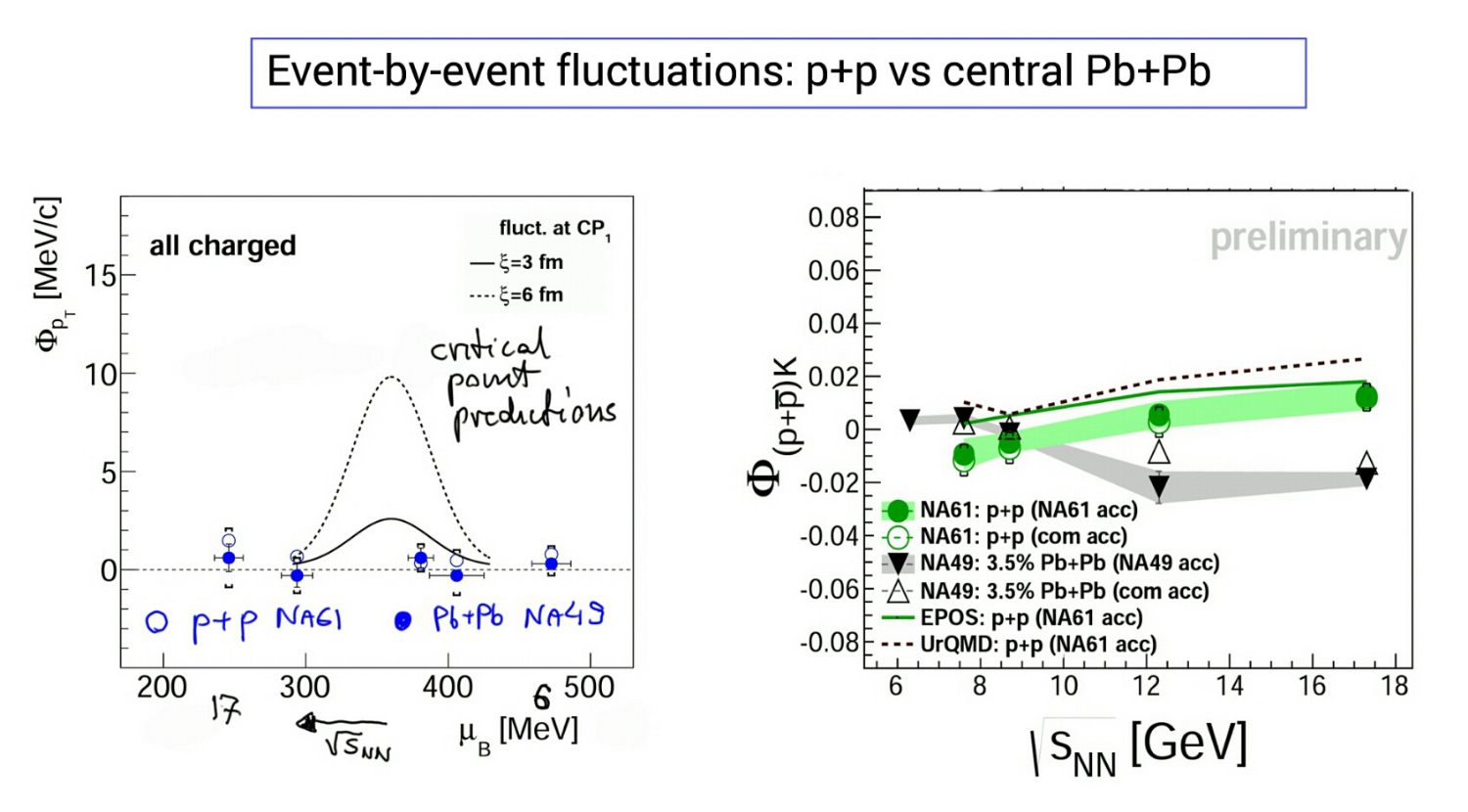}
\caption{ 
Properly normalized transverse momentum (the left plot) and
chemical (the right plot) event-by-event fluctuations
are similar in inelastic p+p interactions and central
Pb+Pb collisions in the SPS energy range.
The largest differences are observed for the relative
[kaon, proton] fluctuations (the right plot).
}
\label{fig:na61_fluct}       
\end{figure}

\clearpage

\newpage
\section{Neutrinos}
\label{sec:Neutrinos}
\begin{figure}[!h]
\centering
\includegraphics[height=4cm,clip]{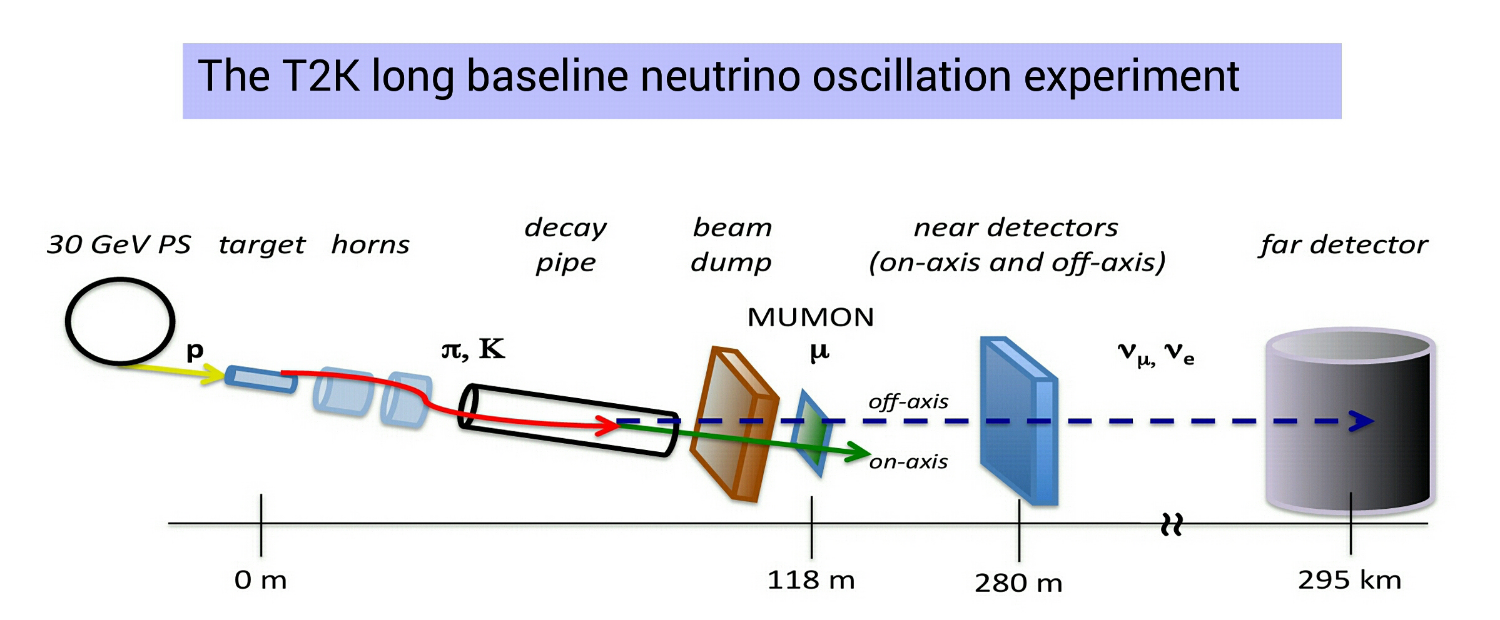}
\caption{ 
The NA61/SHINE measurements for neutrino physics are
motivated by the question how neutrinos change their
identities when flying across Japan.
In order to answer this question the T2K long baseline
neutrino experiment~\cite{Abe:2011ks} sends a neutrino beam from the J-PARC
accelerator at the east coast of Japan to the Kamiokande far detector at the
west coast and compares identified neutrino fluxes at the both locations.
In order to improve information on the initial flux NA61/SHINE
uses a proton beam at the same momentum as in T2K (31~GeV/c)
and measures the flux of hadrons from the T2K replica target
(90 cm long graphite cylinder) and also from
a thin carbon target.
}
\label{fig:t2k}       
\end{figure}

\begin{figure}[!h]
\centering
\includegraphics[height=7cm,clip]{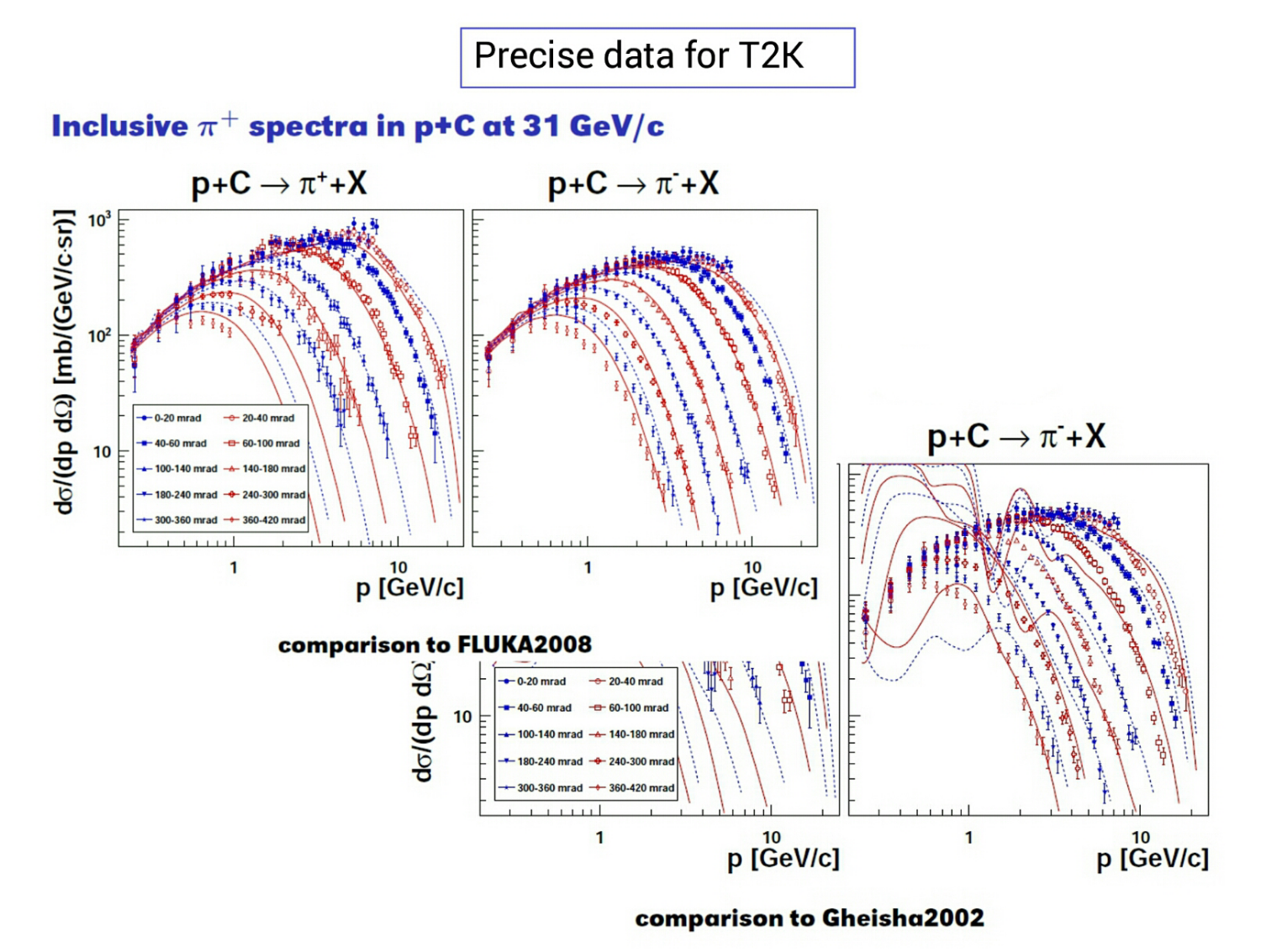}
\caption{ 
The majority of neutrinos in the T2k beam result from decays of 
pions. The NA61/SHINE measurements of pion spectra in p+C interactions
at 31~GeV/c~\cite{Abgrall:2011ae}
are necessary to fit models of primary interactions
in the T2K target~\cite{Korzenev:2013gia}.
}
\label{fig:na61_pC_pionp}       
\end{figure}

\begin{figure}[!h]
\centering
\includegraphics[height=7cm,clip]{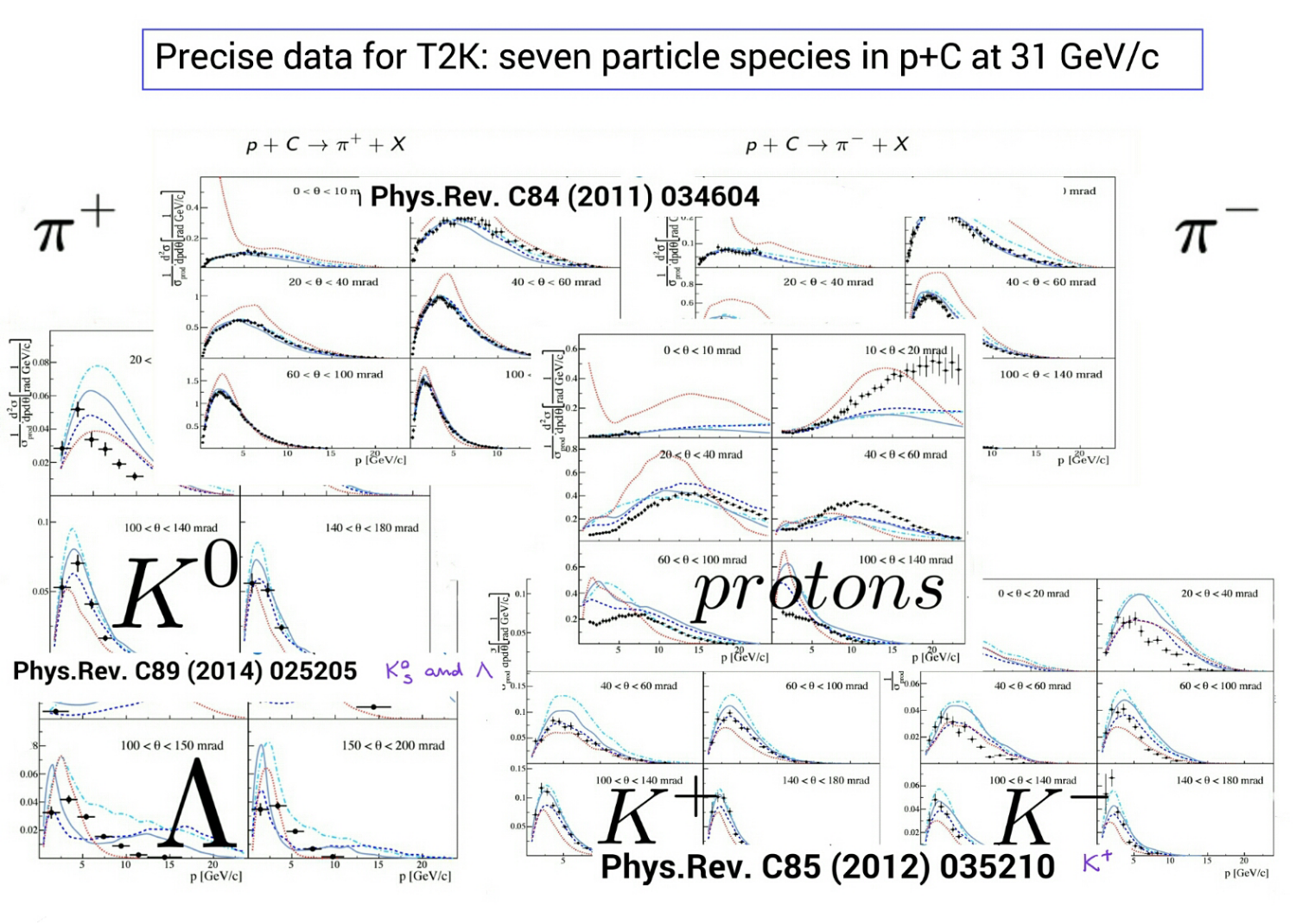}
\caption{ 
In order to further improve the calculations of the initial neutrino
flux measurements of other hadrons abundantly produced in p+C interactions
at 31~GeV/c were performed by NA61/SHINE~\cite{Abgrall:2013wda,
Abgrall:2011ts,Korzenev:2013gia}.
}
\label{fig:na61_pC_all}       
\end{figure}

\begin{figure}[!h]
\centering
\includegraphics[height=7cm,clip]{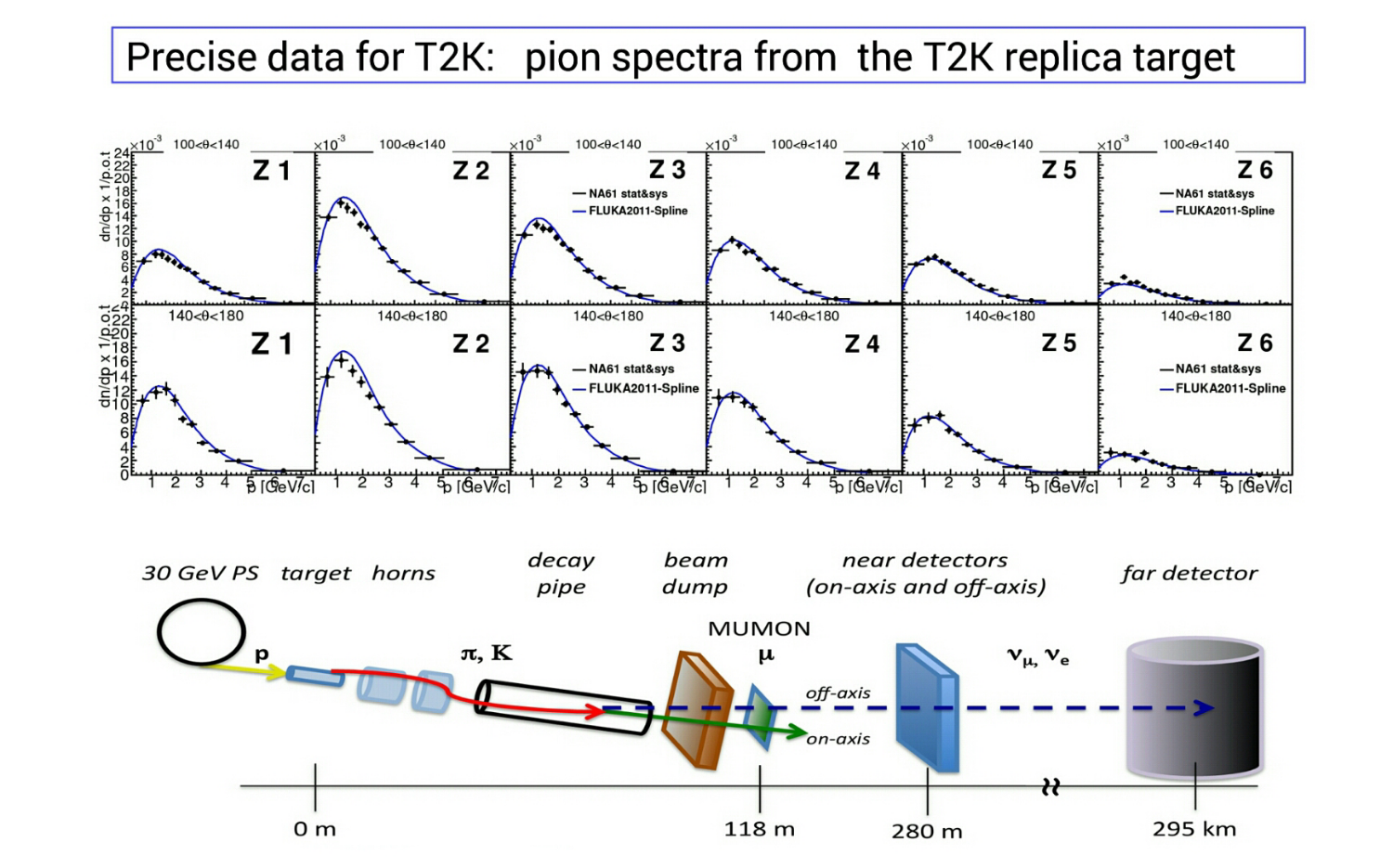}
\caption{ 
In parallel the flux of identified hadrons from the T2K 
replica target was measured~\cite{Abgrall:2012pp}.
The flux is presented separately
for six longitudinal sections of the target (Z1-Z6).
NA61/SHINE plans to perform similar measurements for  
the Femilab neutrino experiments~\cite{add7}.
}
\label{fig:na61_long_target}       
\end{figure}

\clearpage
\newpage
\section{Cosmic rays}
\label{sec:Cosmic_rays}
\begin{figure}[!h]
\centering
\includegraphics[height=6cm,clip]{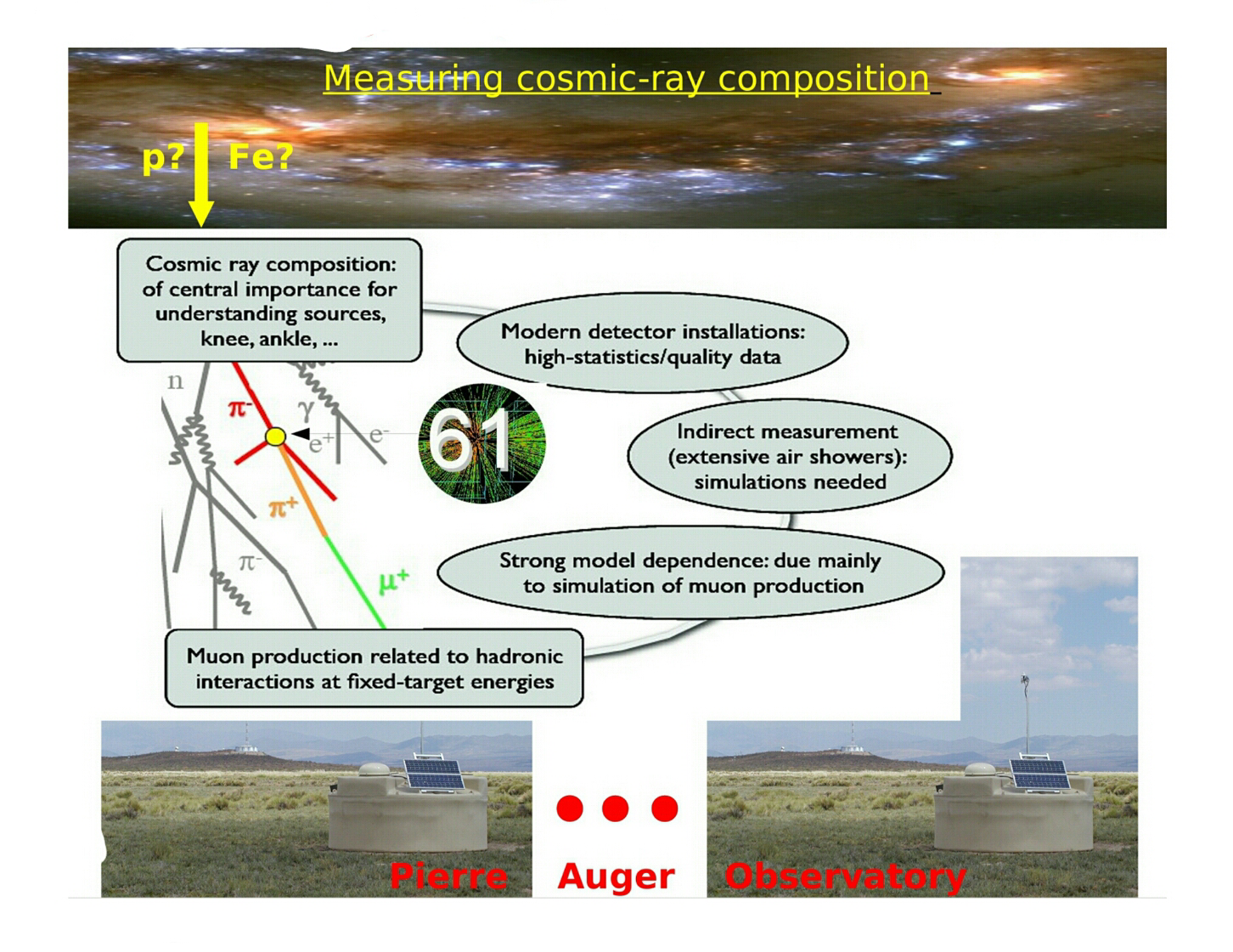}
\caption{ 
The NA61/SHINE measurements for cosmic ray physics are
motivated by the question 
of the origin of very high energy cosmic rays.
In order to find the answer the 
Pierre Auger Observatory~\cite{Abraham:2004dt}
measures properties of extensive air showers (EAS) at the ground level.
These results are then extrapolated back to the primary interactions
of cosmic ray particles with air nuclei aiming to extract
the cosmic ray energy spectrum and composition.  
To reduce model depedence of the extrapolation NA61/SHINE
measured $\pi^-$+C interactions at 158 and 350~GeV/c.
Note that pions are the most abundant hadrons in EAS
and carbon nuclei can well mimic air nuclei.
}
\label{fig:cosmic_rays_intro}       
\end{figure}

\begin{figure}[!h]
\centering
\includegraphics[height=6cm,clip]{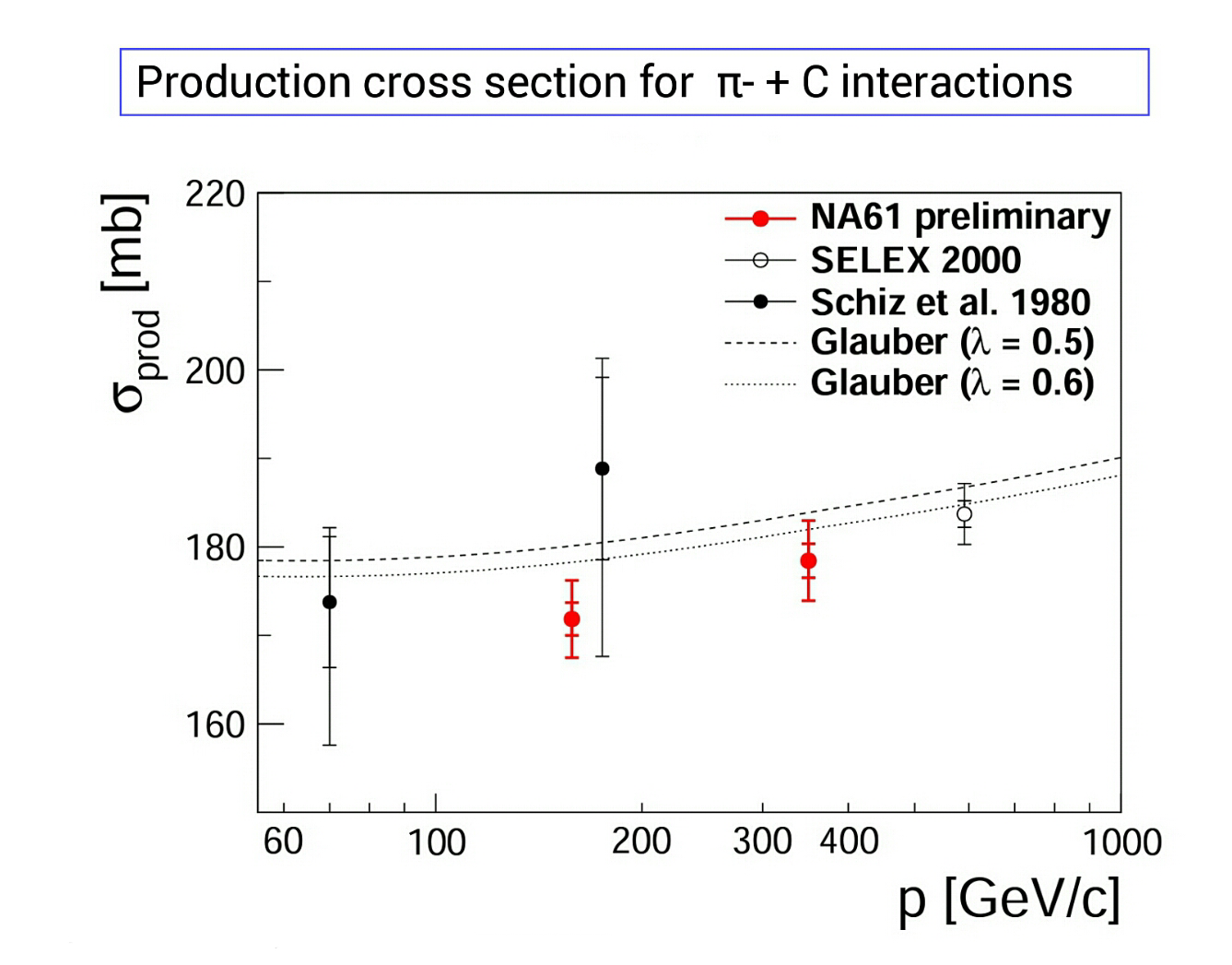}
\caption{ 
Precise measurements of the production cross section for
$\pi^-$+C interactions allow to improve the knowledge about
the probability of
pion-air interactions in EAS~\cite{Abgrall:2014sr}.
}
\label{fig:na61_sigma_piC.png}       
\end{figure}
\begin{figure}[!h]
\centering
\includegraphics[height=8cm,clip]{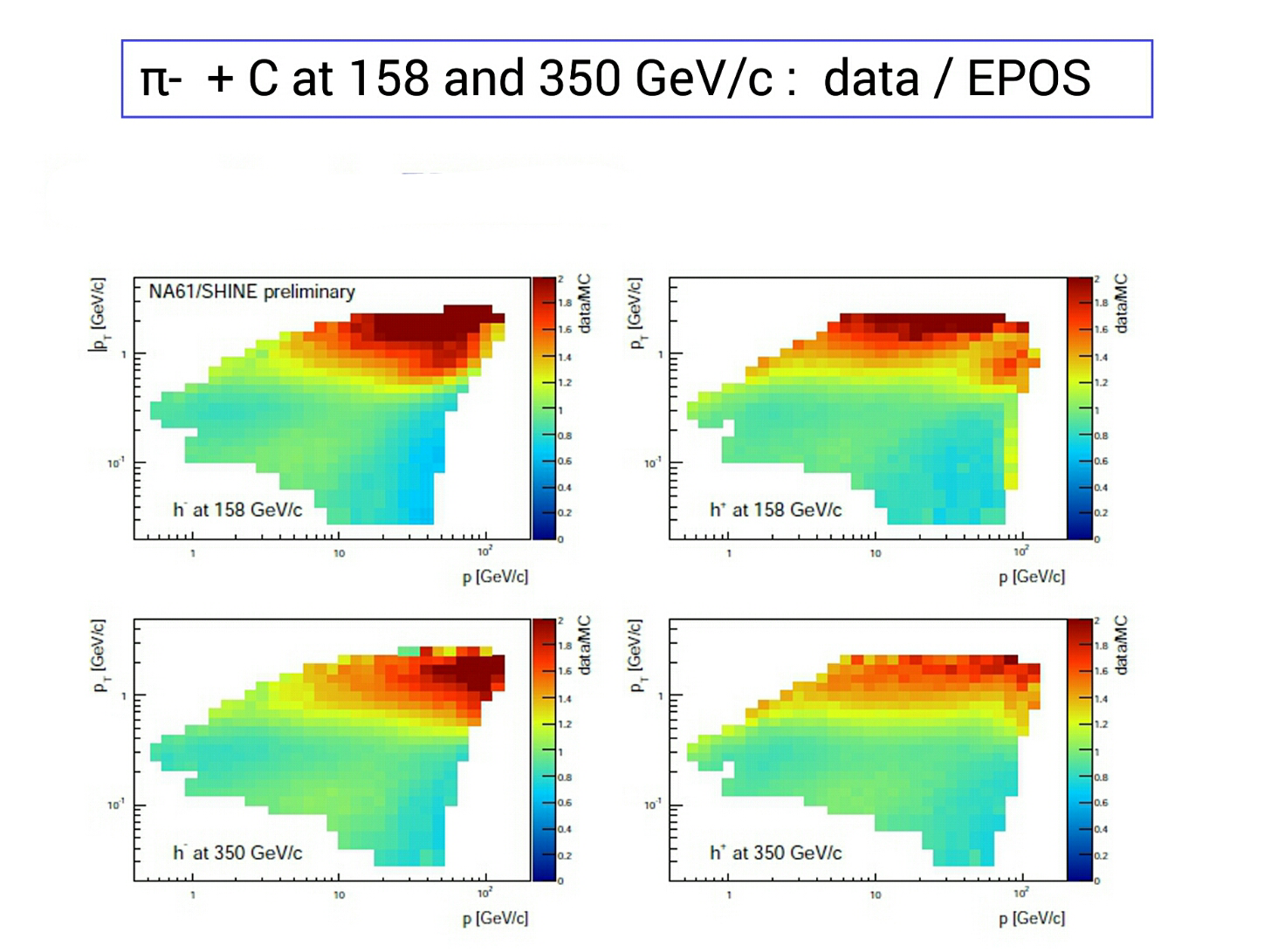}
\caption{ 
Double-differential spectra of charged hadrons produced in
$\pi^-$+C interactions~\cite{M.UngerfortheNA61/SHINE:2013aha}
serve as an input for fitting models
used in EAS simulations.
}
\label{fig:na61_piC_models}       
\end{figure}

\begin{figure}[!h]
\centering
\includegraphics[height=7cm,clip]{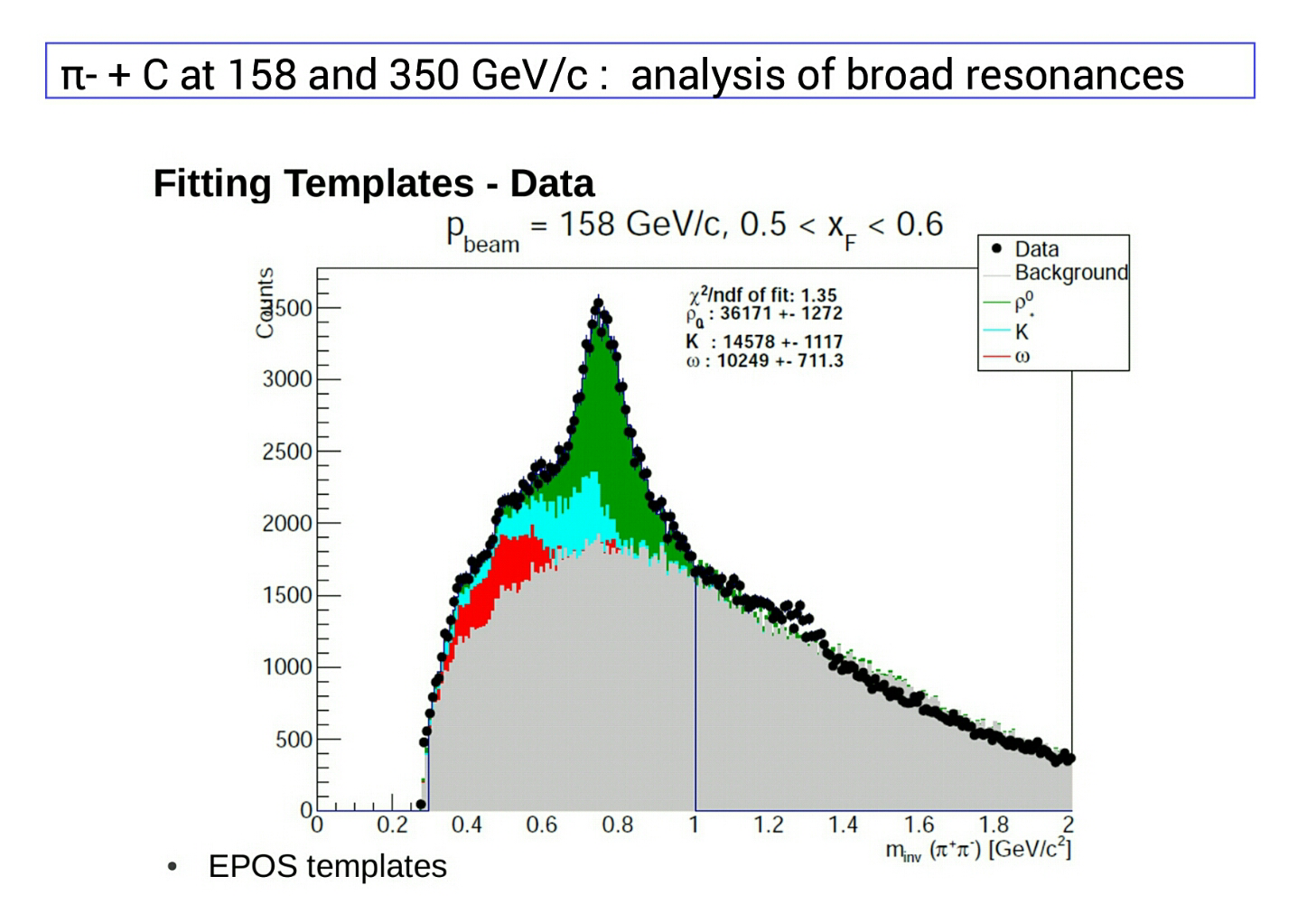}
\caption{ 
Also production of leading $\rho^0$ mesons is important
for the EAS simulations.
Shapes (templates) of invariant mass spectra resulting from decays of various
resonances are calculated using the NA61/SHINE simulation chain.
Then the normalization of the templates is fitted to reproduce the spectrum
measured by the experiment~\cite{Abgrall:2014sr} and to extract the 
resonance yields.
}
\label{fig:na61_piC_rho}       
\end{figure}

%
%
%
\newpage
{\bf Acknowledgements}
\vspace{0.1cm}
This work was supported by
the National Science Centre of Poland (grant
UMO-2012/04/M/ST2/00816) and
the German Research Foundation (grant GA 1480\slash 2-2).

\end{document}